\documentclass[aps,prd,showpacs,amssymb,amsmath]{revtex4}

\usepackage{graphicx}

\begin{document}

\title{Screening masses in neutral two-flavor color superconductor}

\author{Mei Huang}
\email{huang@th.physik.uni-frankfurt.de}
\altaffiliation[on leave
       of absence from ]{%
Physics Department, Tsinghua University,
Beijing 100084, China}%

\author{Igor A. Shovkovy}
\email{shovkovy@th.physik.uni-frankfurt.de}
\altaffiliation[on leave
       of absence from ]{%
       Bogolyubov Institute for Theoretical Physics,
       03143, Kiev, Ukraine}%

\affiliation{%
       Institut f\"{u}r Theoretische Physik,
       J.W. Goethe-Universit\"{a}t,
       D-60054 Frankurt/Main, Germany}%

\date{\today}

\begin{abstract}
The Debye and Meissner screening masses of the gluons and the photon 
in neutral and $\beta$-equilibrated dense two-flavor quark 
matter are calculated. The results are presented in a general form 
that can be used in gapped as well as gapless color superconducting 
phases. The results for the magnetic screening masses indicate that 
the system develops a chromomagnetic instability. Possible consequences 
of the instability are discussed.
\end{abstract}



\maketitle

\section{Introduction}

From the time when the quarks were predicted \cite{GM64}, their nature 
has remained rather elusive. The reason is that direct experimental 
studies of quarks are very limited. Quarks do not exist in vacuum as 
free particles. Under normal conditions, they are always confined 
inside hadrons. The underlying theory of strong interactions quantum  
chromodynamics (QCD) predicts that quarks should become deconfined
at very high temperatures and/or very high densities \cite{qgpN,qgpT}. 
Unfortunately, it is very difficult to achieve sufficiently high 
densities and/or temperatures in laboratory.

Very high temperatures existed in the early Universe during the first 
few microseconds of its evolution \cite{olive}. Nowadays, somewhat 
similar conditions, although at considerably smaller scales and for 
much shorter periods of time, are repeatedly recreated in the so-called 
``little bangs'' at the heavy ion colliders in CERN and at Brookhaven.

Sufficiently high densities may exist in the present Universe inside 
central regions of compact stars. Recently, this possibility attracted 
a lot of attention when it was suggested that various color 
superconducting phases with rather large values of gaps in their 
quasiparticle energy spectra could appear at densities that exist 
inside stars \cite{cs,cfl,weak,weak-cfl}. If this turns out to be true, 
this would be of prime importance. The presence of large gaps in the 
energy spectra can possibly be inferred from a detailed analysis of 
the observational data. This would provide a confirmation of the 
existence of new (quark) states of matter inside compact stars. 

In theoretical studies of dense quark matter, it should be appreciated 
that matter in the bulk of stars is neutral and $\beta$-equilibrated.
Under such conditions, the chemical potentials of different quarks 
should satisfy nontrivial relations. These, in turn, affect the pairing 
dynamics between quarks which is reflected in a specific choice of
the ground state of matter. For example, it was argued in 
Ref.~\cite{no2sc} that a mixture of the two-flavor color 
superconducting (2SC) phase and normal strange quarks is less favorable 
than the color-flavor-locked (CFL) phase after the charge neutrality 
condition is enforced. A similar conclusion was also reached in 
Ref.~\cite{n_steiner}.

Assuming that the constituent medium modified mass of the strange 
quark is large (i.e., larger than the corresponding strange quark 
chemical potential), it was shown recently that neutral two-flavor 
quark matter in $\beta$-equilibrium can have another rather unusual
ground state called the gapless two-flavor color superconductor 
(g2SC) \cite{SH}. While the symmetry in the g2SC ground state is 
the same as that in the conventional 2SC phase, the spectrum of the 
fermionic quasiparticles is different. In particular, two out of four 
gaped quasiparticles of the conventional 2SC phase become gapless in 
the g2SC phase. In addition, the number densities of the pairing 
quarks in the g2SC phase are not equal at zero temperature \cite{SH}. 
For example, the density of red (green) up quarks is different from 
the density of green (red) down quarks.

The existence of gapless color superconducting phases was confirmed 
in Refs.~\cite{GLW,var-appr}, and generalized to nonzero temperatures 
in Refs.~\cite{HS,LZ}. In addition, it was also shown that a gapless 
CFL (gCFL) phase could appear in neutral strange quark matter 
\cite{gCFL,gCFL-long}. At nonzero temperature, the gCFL phase 
and several other new phases (e.g., the so-called dSC and uSC phases)
were studied in Refs.~\cite{dSC,RSR}. If the surface tension is 
sufficiently small, as suggested in Ref.~\cite{RR}, the mixed phase 
composed of the 2SC phase and the normal quark phase will be more 
favored \cite{SHH}. It was also suggested that a non-relativistic 
analogue of gapless superconducting phases could appear in a trapped 
gas of cold fermionic atoms \cite{WilLiu,Deb,LWZ,FGLW}. (Note that an 
alternative ground state for the atomic system, similar to the quark 
mixed phases in Refs.~\cite{neutral_buballa,SHH,RR}, was proposed in 
Ref.~\cite{Bed}.)

While the basic properties of gapless color 
superconducting phases have been established in 
Refs.~\cite{SH,GLW,var-appr,HS,LZ,gCFL,gCFL-long,dSC,RSR},
there is probably much more to be learned about these
phases in the near future. In this paper, we give a detailed 
derivation of the gluon and photon screening masses in neutral 
dense quark matter (the results were briefly presented in 
Ref.~\cite{pi}). We consider the general case of two-flavor 
quark matter, covering both the gapped and the gapless 2SC 
phases. 

The gluon screening properties in the case of the ideal 2SC phase 
(i.e., without a mismatch between the Fermi momenta of different 
quarks) were considered in detail in Refs.~\cite{R-meissner,SWR}. 
Some general features of the gluon screening in the g2SC 
phase were conjectured in Refs.~\cite{HS,kemer} without performing 
the calculation. As we shall see below, the actual results for 
the Debye screening masses are in general agreement with the 
conjecture in Refs.~\cite{HS,kemer}. The Meissner (magnetic) 
screening properties, however, are very different \cite{pi}. 
The calculations in this paper indicate a chromomagnetic 
instability in neutral dense quark matter for a range of 
parameters in the model. As we shall argue in Sec.~\ref{discussion}, 
this instability may lead to a gluon condensation in dense quark 
matter. 

This paper is organized as follows. 
The linear response theory is briefly reviewed in Sec.~\ref{lrt}.
After that, in Sec.~\ref{qp}, we introduce the model and set up the 
main notation. There, we also present the general expression for the 
quark propagator in the color superconducting ground state of neutral 
dense quark matter. 
In Sec.~\ref{gpt}, we briefly discuss the general expression for the
polarization tensor in dense quark matter.
In Sec.~\ref{gpt11}, we study the polarization tensor $\Pi_{\mu\nu}^{AB}$
for $A,B=1,2,3$ and derive the corresponding expressions for the Debye 
and Meissner screening masses. We show that, in accordance with the 
symmetry breaking pattern, there is no Meissner effect (i.e., no Higgs 
mechanism) in this sector of the gauge theory. 
The polarization tensor $\Pi_{\mu\nu}^{AB}$ for $A,B=8,9$ (i.e., the 
8th gluon and the photon) is discussed in Sec. ~\ref{gpt88}.
There we derive the Debye and Meissner screening masses and briefly
discuss their properties. Also, the mixing between the 8th gluon and 
the photon is discussed.
In Sec.~\ref{gpt44}, we study the polarization tensor $\Pi_{\mu\nu}^{AB}$
for $A,B=4,5,6,7$. The results for the Debye and Meissner screening masses 
are presented. As we show, the Debye screening mass is given by a 
rather simple expression that naturally interpolates between the
limits of the normal phase and the ideal 2SC phase. The Meissner mass, 
on the other hand, has an unexpected property. Its value squared is 
negative in a range of parameters, indicating a chromomagnetic 
instability in dense quark matter.
The discussion of the main results is given in Sec.~\ref{discussion}. 
Our findings are summarized in Sec.~\ref{conclusion}.
Several appendices at the end contain useful formulas and some 
details of the calculation. 

\section{Linear response theory and polarization tensor}
\label{lrt}

In order to present a self-contained discussion of the screening 
properties of dense quark matter, we start this paper with a brief 
introduction into the linear response theory and a discussion of
the physical meaning of the polarization tensor in a gauge theory. 
The advanced reader, therefore, may skip this section and go directly 
to Sec.~\ref{qp}.

The response of matter to an external perturbation is the main 
source of knowledge about properties of matter. The linear response 
theory is the simplest framework that is often used to calculate 
such a response. In application to quark matter, 
for example, one studies a response of the system to an external 
source $\mathbf{J}_{\mu}^{A}(x)$. The source is coupled to the 
quantum gauge field. The corresponding interaction part of the 
action reads
\begin{eqnarray}
{\cal S}_{\mathbf{J}} &=& 
\int d^4 x A^{A,\mu}(x) \mathbf{J}_{\mu}^{A}(x) 
\nonumber \\ 
&\equiv & i \int d^4 x \int d^4 y A^{A,\mu}(x) 
\left(D^{-1}\right)_{\mu\nu}^{AB}(x-y) \mathbf{A}^{B,\nu}(y),
\end{eqnarray}
where $\mathbf{A}^{B,\nu}(y)$ is the classical field associated 
with the external source $\mathbf{J}_{\mu}^{A}(x)$, and 
$\left(D^{-1}\right)_{\mu\nu}^{AB}$ is the inverse free gluon 
propagator. Because of the presence of the external source, the 
expectation value of the gauge field becomes nonzero. In the linear 
response theory, it is given by the Kubo's formula \cite{bellac},
\begin{equation}
\langle A^{A}_{\mu} (x) \rangle 
= -i \int d^4 y {\cal D}_{\mu\nu}^{AB}(x-y) \mathbf{J}^{\nu,B}(y),
\end{equation}
where ${\cal D}_{\mu\nu}^{AB}(x-y)$ is the retarded gluon 
propagator. In momentum space, this relation takes the following
form:
\begin{equation}
\langle A^{A}_{\mu} (P) \rangle 
= -i {\cal D}_{\mu\nu}^{AB}(P) \mathbf{J}^{\nu,B}(P),
\end{equation}
where $P^{\mu}\equiv (p^0,\mathbf{p})$ is the energy-momentum 
four-vector. By making use of this result, it is instructive to 
derive an expression for the induced current. It is given by 
\begin{eqnarray}
J_{\mu}^{A,{\rm ind}}(P)
&=&J_{\mu}^{A,{\rm tot}}(P)-\mathbf{J}_{\mu}^{A}(P)\nonumber\\
&=& i \left[\left(D^{-1}\right)_{\mu\nu}^{AB}(P) 
-\left({\cal D}^{-1}\right)_{\mu\nu}^{AB}(P) \right]
\langle A^{B,\nu} (P) \rangle \nonumber\\
&\equiv & \Pi_{\mu\nu}^{AB}(P) 
\langle A^{B,\nu} (P) \rangle ,
\label{J-ind}
\end{eqnarray}
where $\Pi_{\mu\nu}^{AB}(P)$ is the gluon self-energy (or the gluon 
polarization tensor). By definition, this is the one-particle irreducible 
part of the gluon two-point function. The structure of this function is 
constrained by the gauge symmetry. To see this we consider the Slavnov-Taylor 
identity for the full gluon propagator \cite{ST}. 
The explicit form of this identity 
depends on a specific gauge fixing. In the covariant gauge, for example, 
it reads
\begin{equation}
P^{\mu} P^{\nu} i {\cal D}_{\mu \nu}^{AB}(P) =
P^{\mu} P^{\nu} i D_{\mu \nu}^{AB}(P) \equiv
\frac{1}{\lambda},
\label{st1}
\end{equation}
where $\lambda$ is the gauge fixing parameter. In vacuum, where Lorentz 
symmetry is not broken, this relation implies that the gluon 
self-energy is transverse, i.e., $P^{\mu}\Pi_{\mu\nu}^{AB}(P)=0$. 
Because of this constraint, the tensor structure of $\Pi_{\mu\nu}^{AB}(P)$ 
in vacuum is fixed unambiguously,
\begin{eqnarray}
\Pi_{\mu\nu}^{AB}(P)=\Pi^{AB}(P^2)
\left(g_{\mu\nu}-\frac{P_{\mu}P_{\nu}}{P^2}\right),
\end{eqnarray}
where the metric tensor is defined as $g_{\mu\nu}=\mbox{diag}(1,-1,-1,-1)$.
At nonzero temperatures and/or densities, the Lorentz symmetry is broken 
down to its subgroup of spatial rotations $SO(3)$. Then, the tensor 
structure of the gluon self-energy can have a more general form,
\begin{eqnarray}
\Pi^{AB,\mu\nu}(P) &=&
\left(g^{\mu\nu}-u^{\mu}u^{\nu}+\frac{{\mathbf p}^{\mu}{\mathbf p}^{\nu}}
{p^2}\right) H^{AB}
+u^{\mu}u^{\nu} K^{AB} \nonumber\\
&-& \frac{{\mathbf p}^{\mu}{\mathbf p}^{\nu}} {p^2} L^{AB} 
+\frac{u^{\mu}{\mathbf p}^{\nu} +{\mathbf p}^{\mu}u^{\nu}}{p} M^{AB},
\end{eqnarray}
where $u_{\mu}=(1,0,0,0)$ is a time-like four-vector that specifies 
the rest frame of the quark system and ${\mathbf p}^{\mu}= (0,{\mathbf p})$ is
the momentum three-vector with the absolute value $p=|{\mathbf p}|$. 
The component functions $H^{AB}$,
$K^{AB}$, $L^{AB}$, and $M^{AB}$ are functions of $p_0$ and $p$.
Now, the Slavnov-Taylor identity in the covariant gauge takes the form
\begin{equation}
KL + p_0^2 K - p^2 L  + M(M-2p_0 p)=0,
\label{st}
\end{equation}
(for simplicity, the superscripts ``AB'' were omitted here).
It is easy to check that this is less restrictive than the transversality 
condition, $P_{\mu}\Pi^{\mu\nu}(P)=0$, required in vacuum. Indeed, the 
transversality is equivalent to the following set of two relations
between the component functions:
\begin{subequations}
\begin{eqnarray}
L &=& -\frac{p_0^2}{p^2} K \\
M &=& \frac{p_0}{p} K .
\end{eqnarray}
\end{subequations}
While these are sufficient conditions to fulfil the Slavnov-Taylor 
identity in Eq.~(\ref{st}), they are not the necessary conditions in 
a non-Abelian gauge theory when the Lorentz symmetry is broken. (It should 
be emphasized, however, that these are the necessary conditions in Abelian 
gauge theories at nonzero temperatures and/or densities \cite{bellac}.) 

In this paper, we study the polarization tensor $\Pi_{\mu\nu}^{AB}(P)$
in the case of dense quark matter which permits color superconductivity.
In particular, we discuss how the structure of the polarization tensor 
is affected by the (gapless) color superconductivity. 

It is usually said that superconductivity is a result of a gauge symmetry 
breaking. This common misleading statement may suggest that the 
polarization tensor $\Pi_{\mu\nu}^{AB}(P)$ does not need to satisfy the 
Slavnov-Taylor identity (\ref{st1}). In fact, this is not the case because 
a local (gauge) symmetry can never be truly broken \cite{Elitzur}. 

In practice, when doing specific calculations in gauge theories, 
one always breaks the local symmetry by a gauge fixing. As a result, 
it is only a global symmetry that may remain unbroken after the 
gauge choice is made. For example, in the covariant gauge which we 
discussed above, the global color symmetry of the QCD action is 
left unbroken. Then, in a color superconducting phase of quark 
matter, this global symmetry is broken. In such a description, the 
Goldstone theorem requires the appearance of an appropriate number 
of the Nambu-Goldstone bosons (cf., collective excitations in 
Ref.~\cite{Nambu}). Obviously, the appearance of these additional 
degrees of freedom is an artifact of the gauge fixing. Nevertheless, 
their inclusion in the analysis is important in order to insure 
that the general requirements of the gauge symmetry, such as
the Slavnov-Taylor identity (\ref{st1}), are fulfilled \cite{nucl700}. 
Having said this, one should appreciate that the Nambu-Goldstone 
bosons in question are not the physical degrees of freedom. This 
conclusion is easy to reach by noticing that there exist a gauge, 
namely the so-called unitary gauge, in which these bosons can be 
completely eliminated. In a way, their role is similar to the role 
of the Faddeev-Popov ghosts \cite{fp}. While both types of fields 
are unphysical, they are necessary for a consistent description of 
a gauge theory. 

Now let us further discuss the physical meaning of the polarization 
tensor. From a relation similar to that in Eq.~(\ref{J-ind}), it is
clear that, in an Abelian theory such as QED, this tensor is directly 
related to an observable quantity, namely to the induced current. 
The corresponding current is a gauge invariant quantity in an Abelian 
theory. In contrast, the induced current in a non-Abelian theory is 
not a gauge invariant quantity. Then, the physical meaning of 
the polarization tensor is not so clear. In spite of this difficulty, 
we shall use the same interpretation of the polarization tensor in 
quark matter as in an Abelian theory.

In this paper, we study static large-distance, electric and magnetic,
screening properties of quark matter. These describe the response 
of the system to a static perturbation from color/electric charges 
and currents. The static limit means that $p_0=0$. In this case, 
the only nontrivial components of the polarization tensor will be 
$H(p)$ and $K(p)$ which depend only on $p$. Note that
\begin{subequations}
\begin{eqnarray}
\Pi_{00}(0,\mathbf{p}) &=& K(p), \\ 
\Pi_{ij}(0,\mathbf{p}) &=& 
\left(g_{ij} + \frac{{\mathbf p}_i{\mathbf p}_j}{p^2} \right) H(p).
\end{eqnarray}
\label{Pi-K-H}
\end{subequations}
Let us denote the values of the two nontrivial component functions 
in the limit $p \rightarrow 0$ (large distances) as follows:
\begin{subequations}
\begin{eqnarray} \label{DebyeMeissner}
m_D^2 &=& - \lim_{p \rightarrow 0} K(p), \\
m_M^2 &=& - \lim_{p \rightarrow 0} H(p).
\end{eqnarray}
\label{M-K-H}
\end{subequations}
The quantities $m_D$ and $m_M$ are the Debye and Meissner screening 
masses, respectively. It can be shown, see 
for example Ref.~\cite{bellac,kapusta}, that the quantity $m_D$ determines the 
large-distance behavior of the screened potential created by a static 
color/electric charge, i.e., $V(r)\sim \exp(-m_D r)$. By making use of
the analogy with solid state physics systems, we say that a system is a metal 
when $m_D$ is nonzero, and 
it is an insulator when $m_D$ is zero. The quantity $m_M$, when nonzero, 
determines the large-distance fall-off of the (chromo-)magnetic field 
inside a color superconductor, i.e., $B(r)\sim \exp(-m_M r)$. Obviously, 
nonzero $m_M$ is an indication of the Meissner effect. In the normal phase, 
on the other hand, the value of $m_M$ is vanishing. 

\section{Quark propagator}
\label{qp}

In this paper, we continue the study of dense two-flavor quark matter 
constrained by the conditions of the charge neutrality and the 
$\beta$-equilibrium 
\cite{no2sc,n_steiner,SH,GLW,var-appr,HS,LZ,gCFL,gCFL-long,dSC,RSR,n_huang,ruster,bl}.
The use of phenomenological Nambu-Jona-Lasinio (NJL) 
type models has proved to be very convenient in such studies. 
The NJL model can be thought of as a low energy theory 
of QCD in which (massive) gluons, as independent degrees 
of freedom, are integrated out. The gluons could be reintroduced 
back by gauging the color symmetry in the NJL model, 
providing a semirigorous framework for studying the effect of 
the Cooper pairing on the physical properties of gluons. 

In order to study the gluon screening properties in dense quark matter, we 
need to know the quark coupling to the gauge fields. This is determined by 
the quadratic part of the quark Lagrangian density
\begin{equation}
{\cal L}^{(2)} = {\bar \psi} \left( 
i\gamma^{\mu} \partial_{\mu} - m + {\hat \mu} \gamma_0 
+g \gamma^{\mu} A_{\mu}^a T_a 
+ e \gamma^{\mu} A_{\mu} Q \right) \psi, \label{LQA}
\end{equation}
where $T_a$ and $Q$ are the generators of $SU(3)_{\rm c}$ and 
$U(1)_{\rm em}$ gauge groups. The coupling constants of the 
strong interactions and of the electromagnetism are denoted
by $g$ and $e$, respectively. The up and down quark masses are 
assumed to be the same ($m_u=m_d=m$). The quark spinor field has the 
following color-flavor structure:
\begin{equation}
  \label{spinorbasis}
  \psi= \psi_{i \alpha} = \left(
  \begin{array}{c}
     \psi_{ur} \\
     \psi_{ug} \\
     \psi_{ub} \\
     \psi_{dr} \\
     \psi_{dg} \\
     \psi_{db}
  \end{array}
  \right)\, ,
\end{equation}
where $i\in (u,d)$ is the flavor index and $\alpha \in (r,g,b)$ is the color 
index.

In $\beta$-equilibrium, the matrix of chemical potentials in the color-flavor 
space ${\hat \mu}$ is given in terms of the quark chemical potential $\mu$ 
(note that $\mu_B\equiv 3\mu$ is the baryon chemical potential), the chemical
potential for the electrical charge $\mu_e$ and the color chemical potential 
$\mu_8$,  
\begin{eqnarray}
\mu_{ij}^{\alpha\beta} = (\mu \delta_{ij} - \mu_e Q_{ij})\delta^{\alpha\beta}
 + \frac{2}{\sqrt{3}} \mu_8 \delta_{ij} (T_8)^{\alpha\beta}.
\end{eqnarray}
In QCD the color chemical potential $\mu_8$ comes as a result of 
a nonzero expectation value of the 8th gluon field induced by 
the Cooper pairing \cite{DD}. Its absolute value is small because 
it is suppressed parametrically by the quark chemical potential, $\mu_8 \sim 
\Delta^2/\mu$.

The explicit expressions for nontrivial elements of matrix ${\hat \mu}$  
read
\begin{subequations}
\begin{eqnarray}
\mu_{u r} =\mu_{u g} =\mu -\frac{2}{3}\mu_{e} +\frac{1}{3}\mu_{8}, \\
\mu_{d r} =\mu_{d g} =\mu +\frac{1}{3}\mu_{e} +\frac{1}{3}\mu_{8}, \\
\mu_{u b} =\mu -\frac{2}{3}\mu_{e} -\frac{2}{3}\mu_{8}, \\
\mu_{d b} =\mu +\frac{1}{3}\mu_{e} -\frac{2}{3}\mu_{8}. 
\end{eqnarray}
\end{subequations}

To study color superconducting phases, it is convenient to introduce 
the following $8N_cN_f$-component Nambu-Gorkov spinors:
\begin{equation}
  \label{Psi}
  \bar\Psi=\left(\bar\psi,\bar\psi_C\right)\;,\qquad
  \Psi=\left(
  \begin{array}{c}
    \psi \\
    \psi_C
  \end{array}
  \right)\; ,
\end{equation}
where $\psi_C=C\bar\psi^T$ is the charge-conjugate spinor, 
and $C=i \gamma^2\gamma^0$ is the charge-conjugation matrix. 
In this basis, the quadratic part of the quark 
Lagrangian density ${\cal L}^{(2)}$ becomes
\begin{eqnarray}
{\cal L}^{(2)} =   {\bar \Psi} S_0^{-1} \Psi
 + {\bar \Psi}A_\mu^A\hat{\Gamma}^\mu_A \Psi ,
\end{eqnarray}
where the  explicit form of vertex $\hat{\Gamma}^\mu_A$ is
\begin{eqnarray}
\label{vertex}
\hat{\Gamma}_A^\mu\equiv {\rm diag}\left(\Gamma_A^\mu,
\overline{\Gamma}_A^\mu\right)
\equiv \left\{\begin{array}{lll} {\rm diag}(g\,\gamma^\mu T_A,-g\,\gamma^\mu T_A^T) & 
\mbox{for} & \quad A\equiv a=1,\ldots,8 \,\, , \\ \\
{\rm diag}(e\, \gamma^\mu Q,-e\,\gamma^\mu Q) & \mbox{for} & \quad A=9 \quad
\mbox{(photon)} .
\end{array} \right.    
\end{eqnarray}
In momentum space, the inverse free quark propagator $S_0^{-1}$ reads
\begin{equation}
\left[S_0(K)\right]^{-1} = \left(\begin{array}{cc}
\left[G_0^{+}(K)\right]^{-1} & 0 \\
0 & \left[G_0^{-}(K)\right]^{-1}
\end{array}\right).
\end{equation}
The explicit form for the diagonal elements is
\begin{eqnarray}
\left[G_0^{\pm}\right]^{-1} &=& 
 \gamma^0 \left[(k_0-E_{ur}^{\mp})\Lambda^{+}_{k}
+(k_0+E_{ur}^{\pm})\Lambda^{-}_{k} \right]{\cal P}^{(1)}
\nonumber\\
&+&\gamma^0 \left[(k_0-E_{dg}^{\mp})\Lambda^{+}_{k} 
+(k_0+E_{dg}^{\pm})\Lambda^{-}_{k}\right]{\cal P}^{(2)}
\nonumber\\
&+&\gamma^0 \left[(k_0-E_{ub}^{\mp})\Lambda^{+}_{k}
+(k_0+E_{ub}^{\pm})\Lambda^{-}_{k}\right]{\cal P}^{(3)}
\nonumber\\
&+&\gamma^0 \left[(k_0-E_{db}^{\mp})\Lambda^{+}_{k}
+(k_0+E_{db}^{\pm})\Lambda^{-}_{k}\right]{\cal P}^{(4)},\label{G0}
\end{eqnarray}
with the notation $E^{\pm}_{i\alpha}\equiv E_{k}\pm \mu_{i\alpha}$ 
and $E_{k}=\sqrt{k^2+m^2}$. 
The four projectors ${\cal P}^{(I)}$ (with $I=1,\ldots,4$) 
in the 6-dimensional 
color-flavor space are defined as follows:
\begin{subequations}
\begin{eqnarray}
\left( {\cal P}^{(1)}\right)_{ij}^{\alpha\beta}  
&=& (\delta^{\alpha\beta}-\delta^{\alpha b}\delta^{\beta b}) 
\delta_{i u}\delta_{j u},\\
\left( {\cal P}^{(2)}\right)_{ij}^{\alpha\beta}  
&=& (\delta^{\alpha\beta}-\delta^{\alpha b}\delta^{\beta b}) 
\delta_{i d}\delta_{j d},\\
\left( {\cal P}^{(3)}\right)_{ij}^{\alpha\beta}  &=& \delta^{\alpha b}\delta^{\beta b}
\delta_{i u}\delta_{j u} ,\\
\left( {\cal P}^{(4)}\right)_{ij}^{\alpha\beta}  &=& \delta^{\alpha b}\delta^{\beta b}
\delta_{i d}\delta_{j d}.
\end{eqnarray}
\end{subequations}
It is not difficult to see that ${\cal P}^{(1)}$ projects out the red up and 
the green up quarks, and ${\cal P}^{(2)}$ projects out the red down and the 
green down quarks.
The projectors ${\cal P}^{(3)}$ and ${\cal P}^{(4)}$ project out
the blue up and the blue down quarks, respectively. In Eq.~(\ref{G0}),
we also introduced the energy projectors,
\begin{eqnarray}
\Lambda^{\pm}_{k} &=& \frac{1}{2}\left( 1\pm
\gamma^0\frac{\boldsymbol{\gamma}\cdot\mathbf{k}+m}{E_{k}}\right).
\label{Lambda-k}
\end{eqnarray}
These projectors satisfy the following relations \cite{massive_huang}:
\begin{subequations}
\begin{eqnarray}
\gamma^0 \Lambda^{\pm}_{k} \gamma^0 &=& \tilde{\Lambda}^{\mp}_{k},\\
\gamma^5 \Lambda^{\pm}_{k} \gamma^5 &=& \tilde{\Lambda}^{\pm}_{k},
\end{eqnarray}
\end{subequations}
where
\begin{eqnarray}
\tilde{\Lambda}^{\pm}_{k} &=& \frac{1}{2}\left(1\pm
\gamma^0\frac{\boldsymbol{\gamma}\cdot\mathbf{k}-m}{E_{k}} \right)\label{t-Lambda-k}
\end{eqnarray}
is an alternative set of energy projectors. In the chiral limit, the two sets
of projectors in Eqs.~(\ref{Lambda-k}) and (\ref{t-Lambda-k}) coincide.

The full quark propagator in a color superconducting phase takes the following
form:
\begin{equation}
\left[S(K)\right]^{-1} = \left(\begin{array}{cc}
\left[G_0^{+}(K)\right]^{-1} & \Delta^- \\
\Delta^+ & \left[G_0^{-}(K)\right]^{-1}
\end{array}\right),
\label{Spinverse}
\end{equation}
with
\begin{subequations}
\begin{eqnarray}
\Delta^{-} &=& -i \epsilon^{b}\varepsilon\gamma^5 \Delta,\\
\Delta^{+} &\equiv& \gamma^0 \left(\Delta^{-}\right)^{\dagger} \gamma^0
= -i \epsilon^{b}\varepsilon\gamma^5\Delta^{*},
\end{eqnarray}
\end{subequations}
where $\epsilon^b$ is the antisymmetric tensor in the color subspace 
spanned by the red and green colors, while $\varepsilon$ is the 
antisymmetric tensor in the flavor space. The value of the gap 
parameter $\Delta$ is determined from an appropriate gap equation,
while the values of the chemical potentials $\mu_e$ and $\mu_8$ are 
determined from charge neutrality conditions 
\cite{no2sc,n_steiner,SH,var-appr,HS,n_huang,ruster,bl}. 
The explicit form of the gap equation and the neutrality 
conditions is not important for the purposes of this paper.

From Eq.~(\ref{Spinverse}), we obtain 
\begin{equation} 
S = \left(\begin{array}{cc}
G^{+} & \Xi^{-} \\
\Xi^{+} & G^{-}
\end{array}\right), 
\label{quarkpropagator}
\end{equation}
where
\begin{subequations}
\begin{eqnarray}
G^{\pm} &=& 
\left[\left(G_0^{\pm}\right)^{-1} 
-\Delta^{\mp}G_0^{\mp}\Delta^{\pm}\right]^{-1}, \\
\Xi^{\pm}  &=& -G_0^{\mp} \Delta^{\pm} G^{\pm} ,
\end{eqnarray}
\end{subequations}
with the free quark propagators
\begin{eqnarray}
G_0^{\pm} &=&
 \gamma^0 \left[\frac{\tilde{\Lambda}^{+}_{k}}{k_0+E_{ur}^{\pm}}
+\frac{\tilde{\Lambda}^{-}_{k}}{k_0-E_{ur}^{\mp}}\right]{\cal P}^{(1)}
+\gamma^0 \left[\frac{\tilde{\Lambda}^{+}_{k}}{k_0+E_{dg}^{\pm}}
+\frac{\tilde{\Lambda}^{-}_{k}}{k_0-E_{dg}^{\mp}}\right]{\cal P}^{(2)}
\nonumber\\
&+&\gamma^0 \left[\frac{\tilde{\Lambda}^{+}_{k}}{k_0+E_{ub}^{\pm}}
+\frac{\tilde{\Lambda}^{-}_{k}}{k_0-E_{ub}^{\mp}}\right]{\cal P}^{(3)}
+\gamma^0 \left[\frac{\tilde{\Lambda}^{+}_{k}}{k_0+E_{db}^{\pm}}
+ \frac{\tilde{\Lambda}^{-}_{k}}{k_0-E_{db}^{\mp}}\right]{\cal P}^{(4)},
\end{eqnarray}
obtained from Eq.~(\ref{G0}).
By making use of the following relations:
\begin{subequations}
\begin{eqnarray}
\varepsilon {\cal P}^{(1)} \varepsilon &=& - {\cal P}^{(2)}, \qquad 
\epsilon^b {\cal P}^{(1)} \epsilon^b = - {\cal P}^{(1)},\\
\varepsilon {\cal P}^{(2)} \varepsilon &=& - {\cal P}^{(1)}, \qquad 
\epsilon^b {\cal P}^{(2)} \epsilon^b = - {\cal P}^{(2)},\\
\varepsilon {\cal P}^{(3)} \varepsilon &=& - {\cal P}^{(4)}, \qquad 
\epsilon^b {\cal P}^{(3)} \epsilon^b = 0 ,\\
\varepsilon {\cal P}^{(4)} \varepsilon &=& - {\cal P}^{(3)},\qquad  
\epsilon^b {\cal P}^{(4)} \epsilon^b = 0,
\end{eqnarray}
\end{subequations}
one can derive an explicit form of the Nambu-Gorkov components of 
the full propagator,
\begin{subequations}
\begin{eqnarray}
G^{\pm} &=& \sum_{I=1}^4 G^{\pm}_I {\cal P}^{(I)},\\
\Xi^{\pm} &=&\Xi^{\pm}_{12} 
\epsilon^b {\cal P}^{(1)} \varepsilon {\cal P}^{(2)}
+ \Xi^{\pm}_{21} 
\epsilon^b {\cal P}^{(2)} \varepsilon {\cal P}^{(1)}.
\end{eqnarray}
\end{subequations}
The explicit form of the functions $G^{\pm}_I$ and $\Xi^{\pm}_{IJ}$ 
reads
\begin{subequations}
\begin{eqnarray}
\label{quark-propagator}
{\rm G}^{\pm}_1&=&
\frac{k_0-E_{dg}^{\pm}}{(k_0\mp\delta\mu)^2-{E_{\Delta}^{\pm}}^2}\gamma^0
{\tilde \Lambda}_{k}^+ + \frac{k_0+E_{dg}^{\mp}}{(k_0\mp\delta\mu)^2-{E_{\Delta}^{\mp}}^2}
\gamma^0{\tilde \Lambda}_{k}^-,  \\
{\rm G}^{\pm}_2&=&
\frac{k_0-E_{ur}^{\pm}}{(k_0\pm\delta\mu)^2-{E_{\Delta}^{\pm}}^2}\gamma^0
{\tilde \Lambda}_{k}^+ + \frac{k_0+E_{ur}^{\mp}}{(k_0\pm\delta\mu)^2-{E_{\Delta}^{\mp}}^2}
\gamma^0{\tilde \Lambda}_{k}^- , \\
{\rm G}^{\pm}_{3}&=&
\frac{1}{k_0+E_{bu}^{\pm}} \gamma^0{\tilde \Lambda}_{k}^+ + 
\frac{1}{k_0-E_{bu}^{\mp}} \gamma^0{\tilde \Lambda}_{k}^- , \\
{\rm G}^{\pm}_{4}&=&
\frac{1}{k_0+E_{bd}^{\pm}} \gamma^0{\tilde \Lambda}_{k}^+ + 
\frac{1}{k_0-E_{bd}^{\mp}} \gamma^0{\tilde \Lambda}_{k}^- ,
\end{eqnarray}
\label{G_I}
\end{subequations}
and 
\begin{subequations}
\begin{eqnarray}
\Xi^{\pm}_{12} &=& -i \Delta\left(\frac{1}
{(k_0\pm\delta\mu)^2-{E_{\Delta}^{\pm}}^2}\gamma^5{\tilde \Lambda}_{k}^+ +
\frac{1}{(k_0\pm\delta\mu)^2-{E_{\Delta}^{\mp}}^2} \gamma^5 {\tilde \Lambda}_{k}^- \right) ,
 \\
\Xi^{\pm}_{21} &=& -i \Delta\left(\frac{1}
{(k_0\mp\delta\mu)^2-{E_{\Delta}^{\pm}}^2} \gamma^5{\tilde \Lambda}_{k}^+ +\frac{1}
{(k_0\mp\delta\mu)^2-{E_{\Delta}^{\mp}}^2} \gamma^5 {\tilde \Lambda}_{k}^- \right),
\end{eqnarray}
\label{Xi_I}
\end{subequations}
where the following notation was used: 
\begin{subequations}
\begin{eqnarray}
E_{k}^{\pm}&\equiv& E_{k} \pm \bar{\mu},\\
E_{\Delta,k}^{\pm} &\equiv&
\sqrt{(E_{k}^{\pm})^2 +\Delta^2},\\
\bar{\mu} &\equiv&
\frac{\mu_{ur} +\mu_{dg}}{2}
=\frac{\mu_{ug}+\mu_{dr}}{2}
=\mu-\frac{\mu_{e}}{6}+\frac{\mu_{8}}{3}, \label{mu-bar}\\
\delta\mu &\equiv&
 \frac{\mu_{dg}-\mu_{ur}}{2}
=\frac{\mu_{dr}-\mu_{ug}}{2}
=\frac{\mu_{e}}{2}. \label{delta-mu}
\end{eqnarray}
\end{subequations}
In the following sections, we use the full quark propagator 
in Eq.~(\ref{quarkpropagator}) to construct the polarization tensor 
for the gauge fields.

\section{Polarization tensor in dense quark matter}
\label{gpt}

In dense quark matter, screening effects play a very important role 
at length scales larger than the average distance between quarks. 
In the normal phase, for example, the main effects are the Debye 
screening and the Landau damping. These are the properties that 
can be extracted from the behavior of the polarization tensor.
The polarization tensor in dense matter is given by the so-called 
hard dense loop (HDL) approximation \cite{Vija,Manuel}. This 
approximation results from taking into account only the dominant 
one-loop quark contribution in which the internal quark momenta are 
hard (i.e., typical momenta are of order $\mu$). The density of quark 
states with hard momenta is proportional to $\mu^2$ (i.e., the 
density of states at the Fermi surface). Because of this large 
density of states, the quark HDL contribution is large compared 
to the contributions from the gluon and the ghost loops. Therefore, 
the gluon and the ghost contributions are not included in the HDL 
approximation.

In the 2SC/g2SC phase of dense quark matter, the polarization 
tensor is given approximately by the following one-loop expression
\cite{R-meissner,SWR}:
\begin{equation} \label{PiPscfNG}
\Pi^{\mu \nu}_{AB} (P) = \frac{1}{2} \, \frac{T}{V}
\sum_K {\rm Tr}_{\rm D,c,f,NG} \left[ \hat{\Gamma}^\mu_A
{\cal S} (K) \hat{\Gamma}^\nu_B {\cal S}(K-P) \right] ,
\end{equation}
where the trace runs over the Dirac, color, flavor, and 
Nambu-Gokov indices. The gluon part ($A,B=1,\ldots,8$) 
of this polarization tensor reduces to the 
standard HDL result in the normal phase ($\Delta=0$),
\begin{eqnarray}
\Pi^{\mu\nu,ab}_{\rm HDL}(P) &=& \frac{4\alpha_s\mu^2}{\pi}\delta^{ab}
\Bigg\{u^{\mu} u^{\nu} Q\left(\frac{p_0}{p}\right)
-\frac{1}{2}\left(g^{\mu\nu}-u^{\mu} u^{\nu}
+\frac{\mathbf{p}^{\mu}\mathbf{p}^{\nu}}{p^2}\right)
\left[1+\frac{p^2-p_0^2}{p^2}Q\left(\frac{p_0}{p}\right)\right]
\nonumber\\
&&+\frac{\mathbf{p}^{\mu}\mathbf{p}^{\nu}}{p^2}
\frac{p_0^2}{p^2}Q\left(\frac{p_0}{p}\right)
+\frac{p_0}{p}\left(u^{\mu} \frac{\mathbf{p}^{\nu}}{p}
+\frac{\mathbf{p}^{\mu}}{p} u^{\nu}\right)Q\left(\frac{p_0}{p}\right)
\Bigg\},
\label{Pi-HDL}
\end{eqnarray}
where $\alpha_s= g^2/4\pi$, and
\begin{equation}
Q\left(x\right)\equiv -\frac{1}{2} \int_{0}^{1} d \xi
\left(\frac{\xi}{\xi +x -i \varepsilon}+
\frac{\xi}{\xi -x -i \varepsilon}\right)
= \frac{x}{2} \ln \left|\frac{1+x}{1-x}\right|
-1 -i\frac{\pi}{2}|x| \theta(1-x^2).
\end{equation}
In the static limit ($p_0=0$), we obtain $Q(0)=-1$, and the polarization
tensor becomes
\begin{equation}
\Pi^{\mu\nu,ab}_{\rm HDL}(0,\mathbf{p})= -\frac{4\alpha_s\mu^2}{\pi}\delta^{ab}
u^{\mu} u^{\nu}.
\end{equation}
This gluon polarization tensor describes the static screening of
quark color charges at large distances in the normal phase of dense 
quark matter. By comparing with Eqs.~(\ref{Pi-K-H}) and (\ref{M-K-H}), 
we derive the corresponding expression for the Debye screening mass,
\begin{equation}
m_D^2 = \frac{4\alpha_s\mu^2}{\pi}.
\end{equation}
As it should be, the Meissner screening mass is zero in the normal phase.

\section{Gluons with $A=1,2,3$}
\label{gpt11}

In this section, we start with the screening properties of the $A=1,2,3$ gluons.
These are the gluons of the unbroken $SU(2)_c$ subgroup which couple only to the red and 
green quarks. The corresponding expression for the polarization tensor is diagonal,
$\Pi_{AB}^{\mu\nu}(P) \equiv \delta_{AB}\Pi_{11}^{\mu\nu}(P)$.
After performing the traces over the color, flavor and Nambu-Gorkov indices, we 
arrive at 
\begin{eqnarray}
\Pi_{11}^{\mu\nu}(P) &=& \frac{g^2T}{4}\sum_n\int \frac{d^3 {\mathbf k}}{(2\pi)^3}
\mbox{Tr}_{\rm D}  \left[
   \gamma^{\mu} G_{1}^+(K) \gamma^{\nu} G_{1}^+(K')  
    + \gamma^{\mu} G_{1}^-(K) \gamma^{\nu}G_{1}^-(K')\right.\nonumber\\
   &+& \left. 
      \gamma^{\mu} G_{2}^+(K) \gamma^{\nu}G_{2}^+(K')  
    + \gamma^{\mu} G_{2}^-(K) \gamma^{\nu}G_{2}^-(K') \right.\nonumber\\
   &+& \left. 
      \gamma^{\mu}\Xi_{12}^-(K) \gamma^{\nu}\Xi_{21}^+(K') 
    + \gamma^{\mu}\Xi_{12}^+(K) \gamma^{\nu}\Xi_{21}^-(K')\right.\nonumber\\
   &+& \left. 
      \gamma^{\mu}\Xi_{21}^-(K) \gamma^{\nu}\Xi_{12}^+ (K')
    + \gamma^{\mu}\Xi_{21}^+(K) \gamma^{\nu}\Xi_{12}^-(K')\right],
\label{Pi11}
\end{eqnarray}  
where $T$ is the temperature and $K'\equiv K-P$. Here we use the imaginary time
formalism, and the energy integration is replaced by the sum over the fermionic 
Matsubara frequencies $\omega_n\equiv \pi T (2n+1)$.
The explicit expressions for the components $G^{\pm}_{I}$ and 
$\Xi^{\pm}_{IJ}$ of the quark propagator are given in Eqs.~(\ref{G_I}) and (\ref{Xi_I}).
After the summation over the Matsubara frequencies, we obtain the result in the 
following form:
\begin{eqnarray}
\Pi_{11}^{\mu\nu}(P) &=& \pi\alpha_s\int \frac{d^3 {\mathbf k}}{(2\pi)^3}
  \left[
   \left(C_{++}^{11} + C_{++}^{22}\right){\cal T}_{++}^{\mu\nu}
  + \left(C_{--}^{11} + C_{--}^{22}\right){\cal T}_{--}^{\mu\nu}\right.\nonumber\\
   &+& \left. 
   \left(C_{+-}^{11} + C_{+-}^{22}\right){\cal T}_{+-}^{\mu\nu}
  + \left(C_{-+}^{11} + C_{-+}^{22}\right){\cal T}_{-+}^{\mu\nu}\right.\nonumber\\
   &+& \left. 
   \left(C_{++}^{12} + C_{++}^{21}\right){\cal U}_{++}^{\mu\nu}
  + \left(C_{--}^{12} + C_{--}^{21}\right){\cal U}_{--}^{\mu\nu}\right.\nonumber\\
   &+& \left. 
   \left(C_{+-}^{12} + C_{+-}^{21}\right){\cal U}_{+-}^{\mu\nu}
  + \left(C_{-+}^{12} + C_{-+}^{21}\right){\cal U}_{-+}^{\mu\nu} \right].
\label{Pi11-n}
\end{eqnarray} 
In this expression, we introduced the following notation for the two types 
of Dirac traces:
\begin{subequations}
\begin{eqnarray}
{\cal T}_{e_1e_2}^{\mu\nu}& = &{\rm Tr}_{\rm D}[\gamma^{\mu}\gamma^0{\tilde \Lambda}_{k}^{e_1}
    \gamma^{\nu}\gamma^0{\tilde \Lambda}_{k'}^{e_2}],  \\
{\cal U}_{e_1e_2}^{\mu\nu}& = & {\rm Tr}_{\rm D}[\gamma^{\mu}\gamma^5{\tilde \Lambda}_{k}^{e_1}
    \gamma^{\nu}\gamma^5{\tilde \Lambda}_{k'}^{e_2}],
\end{eqnarray} 
\label{D-traces}
\end{subequations}
with $e_1, e_2 = \pm$. To leading order in $1/\mu$, the results for these traces 
are given in Eqs. (\ref{tr2++})--(\ref{tr1+-}) in Appendix~\ref{TU-matrix}.
The expressions for the coefficient functions $C_{\pm\pm}^{IJ}$ at zero and nonzero 
temperatures are given in Appendix~\ref{Cof-TU11}. 

We write the integral over the three-momentum in Eq. (\ref{Pi11-n}) as
\begin{eqnarray}
\int d^3 {\mathbf k}=\int \frac{d k}{4 \pi^2} k^2 \int_{-1}^1 
d \xi \int_0^{2\pi} \frac{d \phi}{2 \pi},
\end{eqnarray}
where $\phi$ is the polar angle and $\xi$ is the cosine of the angle between 
the three-momenta ${\mathbf k}$ and ${\mathbf p}$. After performing the integral
over the polar angle $\phi$, see Eqs.~(\ref{phi-int1}) and (\ref{phi-int2}), 
the corresponding traces in the integrand can be replaced by the following angular
averaged expressions:
\begin{subequations}
\begin{eqnarray}
{\cal T}^{\mu\nu}_{\pm\pm} &\to &
2u^{\mu}u^{\nu}
\mp 2\xi \frac{u^{\mu} {\mathbf p}^{\nu}+{\mathbf p}^{\mu}u^{\nu}}{p}
-(1-\xi^2)( g^{\mu\nu} - u^{\mu}u^{\nu} )
-(1-3\xi^2)\frac{{\mathbf p}^{\mu}{\mathbf p}^{\nu}}{p^{2}},
\label{22++} \\
{\cal T}^{\mu\nu}_{\pm\mp}&\to &
-(1+\xi^2)( g^{\mu\nu} - u^{\mu}u^{\nu} )
+(1-3\xi^2)\frac{{\mathbf p}^{\mu}{\mathbf p}^{\nu}}{p^{2}},
\label{22+-} \\
{\cal U}^{\mu\nu}_{\pm\pm} &\to&
-(1+\xi^2)( g^{\mu\nu} - u^{\mu}u^{\nu} )
+(1-3\xi^2)\frac{{\mathbf p}^{\mu}{\mathbf p}^{\nu}}{p^{2}},
\label{11++}\\
{\cal U}^{\mu\nu}_{\pm\mp} &\to &
-2 u^{\mu}u^{\nu}
-(1-\xi^2)( g^{\mu\nu} - u^{\mu}u^{\nu} )
-(1-3\xi^2)\frac{{\mathbf p}^{\mu}{\mathbf p}^{\nu}}{p^{2}}
\mp 2\xi \frac{u^{\mu} {\mathbf p}^{\nu}-{\mathbf p}^{\mu}u^{\nu}}{p}.
\label{11+-}
\end{eqnarray}
\label{traces-xi}
\end{subequations}
Now we would like to note that there are two different types of coefficient 
functions in Eq.~(\ref{Pi11-n}). The coefficients $C_{\pm\pm}^{11}$, 
$C_{\pm\pm}^{22}$,  $C_{\pm\mp}^{12}$ and $C_{\pm\mp}^{21}$ originate from 
particle-hole and antiparticle-antiparticle loops. In the leading order 
approximation, we drop the antiparticle-antiparticle contributions to the 
polarization tensor. These are suppressed by an inverse power of the quark chemical 
potential. In addition, in the static long wavelength limit ($p_0=0$ and $p\to 0$),
the particle-hole contributions simplify considerably. The approximate expressions
read
\begin{subequations}
\begin{eqnarray}
C_{\pm\pm}^{11} & \simeq & 
-\frac{\Delta^2}{4 (E_{\Delta, k}^{-})^3}[1-\theta(-E_{\Delta, k}^- + \delta\mu)] 
- \delta(-E_{\Delta, k}^- + \delta\mu) 
\frac{ (E_{\Delta, k}^-)^2+( E_k^-)^2-2 E_{\Delta, k}^-  E_k^- }
{4 (E_{\Delta, k}^-)^2}, \\
C_{\pm\pm}^{22} & \simeq & 
-\frac{\Delta^2}{4 (E_{\Delta, k}^{-})^3}[1-\theta(-E_{\Delta, k}^- + \delta\mu)] 
- \delta(-E_{\Delta, k}^- + \delta\mu) \frac{(E_{\Delta, k}^-)^2+( E_{k}^-)^2+
2 E_{\Delta, k}^-  E_k^- }{4 (E_{\Delta, k}^-)^2}, \\
C_{\pm\mp}^{12}& = & C_{\pm\mp}^{21}  \simeq
 -\frac{\Delta^2}{4 (E_{\Delta, k}^{-})^3}[1-\theta(-E_{\Delta, k}^- + \delta\mu)] 
+ \delta(-E_{\Delta, k}^- + \delta\mu) \frac{\Delta^2}{4 (E_{\Delta, k}^-)^2}.
\end{eqnarray}
\label{p-h}
\end{subequations}
Here we neglected the corrections due to nonzero quark masses. This is 
justified if the shift of the quark Fermi momenta due to such 
masses is small, i.e., $m^2/\mu \ll \delta\mu$. We also used the 
following relations valid when $p\to 0$:
\begin{subequations}
\begin{eqnarray} \label{limit}
k' & = &|{\mathbf k} -{\mathbf p}| \simeq k - p \xi  , \\
 E_{k'}^{\pm} & \simeq  &  E_{k}^{\pm} - \frac{k}{E_k} p \xi,  \\
E_{\Delta,k'}^{\pm} & \simeq  & E_{\Delta,k}^{\pm} - 
  \frac{k}{E_k}\frac{ E_k^{\pm}}{E_{\Delta,k}^{\pm}} p \xi, \\
\theta(-E_{\Delta,k'}^- + \delta\mu) & \simeq  & 
\theta(-E_{\Delta,k}^- + \delta\mu) + \delta(-E_{\Delta,k}^- + \delta\mu )\frac{k}{E_k}
 \frac{ E_k^-}{E_{\Delta,k}^-} p \xi.
\end{eqnarray}
\end{subequations}
As is easy to check, the terms with the $\theta$- and $\delta$-functions in 
Eq.~(\ref{p-h}) contribute only in the g2SC phase when $\Delta<\delta\mu$ 
(this also includes the normal phase as a limiting case with $\Delta=0$).

Now, the coefficients $C_{\pm\mp}^{11}$, $C_{\pm\mp}^{22}$,  
$C_{\pm\pm}^{12}$ and $C_{\pm\pm}^{21}$ are made of particle-antiparticle 
loops only. Their contributions to the polarization tensor are not negligible 
although they are formally suppressed by an inverse power of the quark 
chemical potential. In fact, the corresponding contributions are 
ultraviolet divergent. The divergences are removed by subtractions of 
the corresponding vacuum terms. One can also check that a nonzero gap 
parameter ($\Delta\ll \mu$) leads only to a small correction to these 
particle-antiparticle contributions. Thus, we approximate the corresponding 
coefficient functions as follows:
\begin{subequations}
\begin{eqnarray}
C_{\pm\mp}^{11} & \simeq & C_{\pm\mp}^{22} \simeq 
-\frac{|k - \bar\mu | + k - \bar\mu }
{ |k - \bar\mu | (k + \bar\mu + |k - \bar\mu |)} +\frac{1}{k}, \\
C_{\pm\pm}^{12}& = & C_{\pm\pm}^{21} \simeq 0.
\end{eqnarray} 
\label{p-a}
\end{subequations}
While the traces in Eq.~(\ref{traces-xi}) depend on the 
angular coordinate $\xi$, the approximate coefficient functions 
$C_{\pm\pm}^{IJ}$ in Eqs.~(\ref{p-h}) and (\ref{p-a}) do not. 
Therefore, the corresponding angular integration in Eq.~(\ref{Pi11-n}) 
can be easily performed,
\begin{subequations}
\begin{eqnarray}
\int_{-1}^{1} d \xi ~{\cal T}^{\mu\nu}_{\pm\pm} &= &
4u^{\mu}u^{\nu}
-\frac{4}{3}( g^{\mu\nu} - u^{\mu}u^{\nu} ),
\label{222++} \\
\int_{-1}^{1} d \xi ~{\cal T}^{\mu\nu}_{\pm\mp}&=&
-\frac{8}{3}( g^{\mu\nu} - u^{\mu}u^{\nu} ),
\label{222+-}\\
\int_{-1}^{1} d \xi ~{\cal U}^{\mu\nu}_{\pm\pm} &=&
- \frac{8}{3} ( g^{\mu\nu} - u^{\mu}u^{\nu} ),
\label{111++}\\
\int_{-1}^{1} d \xi ~{\cal U}^{\mu\nu}_{\pm\mp} &=&
-4 u^{\mu}u^{\nu}
- \frac{4}{3}( g^{\mu\nu} - u^{\mu}u^{\nu} ).
\label{111+-}
\end{eqnarray}
\label{tu-integral}
\end{subequations}
The results on the right hand side are independent of the momentum $p$. 
Therefore, in order to derive the final expression for the polarization 
tensor, we need to calculate only the following momentum integrals:
\begin{subequations}
\begin{eqnarray}
\int d k k^2 C_{\pm\pm}^{11,22} 
& \simeq & - \frac{1}{2} ~ {\bar \mu}^2 \left[1+
\frac{\delta\mu\, \theta( \delta\mu-\Delta)}{\sqrt{{\delta\mu}^2-\Delta^2}} 
\right], 
\label{ph-t}\\ 
\int dk k^2  C_{\pm\mp}^{12,21}
& \simeq &  - \frac{1}{2} ~ {\bar \mu}^2 \left[1- 
\frac{\delta\mu\, \theta( \delta\mu-\Delta)}{\sqrt{{\delta\mu}^2-\Delta^2}} 
\right]  
\label{ph-u},\\
\int d k k^2 C_{\pm\mp}^{11,22}  & \simeq & \frac{1}{2}{\bar \mu}^2, 
\label{panp-t} \\ 
\int d k k^2 C_{\pm\pm}^{12,21}  & \simeq & 0 . 
\label{panp-u}
\end{eqnarray}
\label{c-integral}
\end{subequations}
The details of the calculation are given in Appendix~\ref{Int-11}.
Note that, in the 2SC phase when $\delta\mu < \Delta$, the contributions 
from the normal ($C^{11,22}_{\pm\pm}$) and abnormal ($C^{12,21}_{\pm\mp}$)
quark loops are equal, see Eqs.~(\ref{ph-t}) and (\ref{ph-u}). The 
additional contributions in the g2SC phase have equal absolute values 
but differ in sign. As for the particle-antiparticle quark loop, only 
the normal ($C^{11,22}_{\pm\mp}$) contribution is nonvanishing.

By substituting the results in Eqs. (\ref{tu-integral}) and 
(\ref{c-integral}) into Eq.~(\ref{Pi11-n}), we derive the following 
expression for the polarization tensor $\Pi_{11}^{\mu\nu}$: 
\begin{eqnarray}
\Pi_{11}^{\mu\nu}(0) &\simeq & \frac{4\alpha_s}{3\pi} \int d k k^2
  \left[3 \left(C_{++}^{11} - C_{+-}^{12} \right)u^{\mu}u^{\nu}
  -\left(C_{++}^{11}+2C_{+-}^{11}+C_{+-}^{12}\right)
   \left(g^{\mu\nu}-u^{\mu}u^{\nu}\right) \right] \nonumber\\
&\simeq & -\frac{4\alpha_s \bar\mu^2}{\pi} \frac{\delta\mu}
{\sqrt{{\delta\mu}^2-\Delta^2}} \theta(\delta\mu-\Delta) u^{\mu}u^{\nu}.
\label{scr11}
\end{eqnarray}
Only the $00$-component of this polarization tensor is nonzero.
This component determines the Debye screening mass,
\begin{eqnarray}
m_{D,11}^2\equiv -\Pi^{00}_{11}(0) \simeq  
\frac{4\alpha_s \bar\mu^2}{\pi} \frac{\delta\mu}
{\sqrt{{\delta\mu}^2-\Delta^2}} \theta(\delta\mu-\Delta) .
\label{m_D_1}
\end{eqnarray}
In the 2SC phase ($\Delta>\delta\mu$), where there are no gapless 
excitations charged with respect to the unbroken $SU(2)_c$ subgroup, 
this screening mass is identically zero \cite{R-meissner}. In contrast, 
it is nonzero in the g2SC phase ($\Delta<\delta\mu$). In fact, its value 
is proportional to the density of the gapless quasiparticle states.

As for the Meissner screening mass, its value vanishes in the 2SC 
phase as well as in the g2SC phase,
\begin{eqnarray}
m_{M,11}^2 & \equiv & -\frac{1}{2}\lim_{p\to 0} \left(g_{ij} + 
\frac{\mathbf{p}_i\mathbf{p}_j}{p^2} \right) 
\Pi_{11}^{ij}(0,\mathbf{p}) \simeq 0 .
\end{eqnarray}


\section{8th gluon, photon and the gluon-photon mixing}
\label{gpt88}

In the 2SC/g2SC phase, the diquark condensate breaks the gauge symmetry 
$SU(3)_c \times U(1)_{\rm em}$ down to the $SU(2)_c\times \tilde{U}(1)_{\rm em}$ 
subgroup. The structure of the condensate in the ground is such that the 
8th gluon and the photon mix. The new medium fields are given by the following 
linear combinations of the vacuum fields \cite{mag}:
\begin{subequations}
\begin{eqnarray}
\tilde{A}^8_\mu = A^8_\mu \cos\varphi + A^\gamma_\mu \sin\varphi 
\label{tilde-A},\\
\tilde{A}^\gamma_\mu = A^\gamma_\mu \cos\varphi  - A^8_\mu \sin\varphi ,
\label{tilde-gamma}
\end{eqnarray}
\label{tilde-A-gamma}
\end{subequations}
where the mixing angle $\varphi$ is determined from the structure of the
condensate by using simple group-theoretical arguments,
\begin{subequations}
\begin{eqnarray}
\cos\varphi & = &\sqrt{\frac{3\alpha_s}{3\alpha_s+\alpha}}, \\
\sin\varphi & = &\sqrt{\frac{\alpha}{3\alpha_s+\alpha}}.
\end{eqnarray}
\end{subequations}
The generator of the medium (low-energy) electromagnetism $\tilde{U}(1)_{\rm em}$ 
is defined as follows: $\tilde{Q}=Q-\frac{1}{\sqrt{3}} T_8$. 

The 8th gluon can probe the Cooper-paired red and green quarks, as 
well as the unpaired blue quarks. After the traces over the color, the 
flavor and the Nambu-Gorkov indices are performed, the polarization 
tensor for the 8th gluon can be expressed as 
\begin{subequations}
\begin{eqnarray}
\Pi_{88}^{\mu\nu}(P) &=& \frac{1}{3}{\tilde \Pi}_{88}^{\mu\nu}(P)+  \frac{2}{3}
\Pi_{88,b}^{\mu\nu}(P),  \\ \label{Pi88-2}
{\tilde \Pi}_{88}^{\mu\nu}(P)
&=& \frac{g^2T}{4}\sum_n\int \frac{d^3 {\mathbf k}}{(2\pi)^3}
\mbox{Tr}_{\rm D}  \left[
   \gamma^{\mu} G_{1}^+(K) \gamma^{\nu} G_{1}^+(K')  
    + \gamma^{\mu} G_{1}^-(K) \gamma^{\nu}G_{1}^-(K')\right.\nonumber\\
   &+& \left. 
      \gamma^{\mu} G_{2}^+(K) \gamma^{\nu}G_{2}^+(K')  
    + \gamma^{\mu} G_{2}^-(K) \gamma^{\nu}G_{2}^-(K') \right.\nonumber\\
   &-& \left. 
      \gamma^{\mu}\Xi_{12}^-(K) \gamma^{\nu}\Xi_{21}^+(K') 
    - \gamma^{\mu}\Xi_{12}^+(K) \gamma^{\nu}\Xi_{21}^-(K')\right.\nonumber\\
   &-& \left. 
      \gamma^{\mu}\Xi_{21}^-(K) \gamma^{\nu}\Xi_{12}^+ (K')
    - \gamma^{\mu}\Xi_{21}^+(K) \gamma^{\nu}\Xi_{12}^-(K')\right],
\label{Pi88} \\ 
\Pi_{88,b}^{\mu\nu}(P) &=&  \frac{g^2T}{4}\sum_n\int \frac{d^3 {\mathbf k}}{(2\pi)^3}
\mbox{Tr}_{\rm D}  \left[ \gamma^{\mu} G_{3}^+(K) \gamma^{\nu}
G_{3}^+(K')  + \gamma^{\mu} G_{3}^-(K) \gamma^{\nu}G_{3}^-(K')
\right.\nonumber\\
   &+& \left. \gamma^{\mu} G_{4}^+(K) 
\gamma^{\nu}G_{4}^+(K')  + \gamma^{\mu} G_{4}^-(K) \gamma^{\nu}G_{4}^-(K')
\right].
\label{Pi88b}
\end{eqnarray}
\end{subequations}
Note that the contribution ${\tilde \Pi}_{88}^{\mu\nu}$ of paired quarks,
up to a sign in front of the contribution from the abnormal quark loops, 
is the same as the expression for $\Pi_{11}^{\mu\nu}$ in Eq.~(\ref{Pi11}). 
Therefore, by following the same steps as in the previous section, we 
easily derive the result
\begin{eqnarray}
\tilde{\Pi}_{88}^{\mu\nu}(0) &\simeq & \frac{4\alpha_s}{3\pi} \int d k k^2
  \left[3 \left(C_{++}^{11} + C_{+-}^{12} \right)u^{\mu}u^{\nu}
  -\left(C_{++}^{11}+2C_{+-}^{11}-C_{+-}^{12}\right)
   \left(g^{\mu\nu}-u^{\mu}u^{\nu}\right) \right] \nonumber\\
&\simeq & -\frac{4\alpha_s \bar\mu^2}{\pi} u^{\mu}u^{\nu}
-\frac{4\alpha_s \bar\mu^2}{3\pi} \left[ 1-
\frac{\delta\mu\, \theta(\delta\mu-\Delta)}{\sqrt{{\delta\mu}^2-\Delta^2}} 
\right]
\left( g^{\mu\nu}-u^{\mu}u^{\nu}\right),
\label{scr88-1}
\end{eqnarray}
[c.f. Eq.~(\ref{scr11})]. The unpaired blue quarks give the standard 
normal phase HDL contribution 
$\Pi_{88,b}^{\mu\nu}(P)\equiv \Pi_{\rm HDL}^{\mu\nu}(P)$, 
see Eq.~(\ref{Pi-HDL}).
In the static ($p_0=0$) long-wavelength ($p\to 0$) limit, 
this leads to the following result:
\begin{eqnarray}
\Pi_{88,b}^{\mu\nu}(0) = - \frac{4 \alpha_s}{\pi} {\bar\mu}^2
u^{\mu}u^{\nu}.
\end{eqnarray}
 Thus, the final result for the polarization tensor of the 8th 
gluon reads
\begin{eqnarray}
\Pi_{88}^{\mu\nu}(0)  
&\simeq & -\frac{4\alpha_s \bar\mu^2}{\pi} u^{\mu}u^{\nu}
-\frac{4\alpha_s \bar\mu^2}{9\pi} \left[ 1-
\frac{\delta\mu\, \theta(\delta\mu-\Delta)}{\sqrt{{\delta\mu}^2-\Delta^2}} 
\right]
\left( g^{\mu\nu}-u^{\mu}u^{\nu}\right).
\label{scr88}
\end{eqnarray}
Because of the symmetry considerations, one should expect a nontrivial
gluon-photon mixing in the ground state. So, we proceed to the calculation of 
the photon polarization tensor. The general expression for this tensor is
\begin{subequations}
\begin{eqnarray}
\Pi_{99}^{\mu\nu}(P) &=& \frac{1}{2}{\tilde \Pi}_{99}^{\mu\nu}(P)+ 
\frac{1}{2}\Pi_{99,b}^{\mu\nu}(P), \label{Pi99-1} \\ 
{\tilde \Pi}_{99}^{\mu\nu}(P)
&=& \frac{2e^2T}{9}\sum_n\int \frac{d^3 {\mathbf k}}{(2\pi)^3}
\mbox{Tr}_{\rm D}  \left\{
   4 \left[\gamma^{\mu} G_{1}^+(K) \gamma^{\nu} G_{1}^+(K')  
    + \gamma^{\mu} G_{1}^-(K) \gamma^{\nu}G_{1}^-(K')\right]
\right.\nonumber\\
   &+& \left. 
      \gamma^{\mu} G_{2}^+(K) \gamma^{\nu}G_{2}^+(K')  
    + \gamma^{\mu} G_{2}^-(K) \gamma^{\nu}G_{2}^-(K') \right.\nonumber\\
   &+& \left. 
      2\left[\gamma^{\mu}\Xi_{12}^-(K) \gamma^{\nu}\Xi_{21}^+(K') 
    + \gamma^{\mu}\Xi_{12}^+(K) \gamma^{\nu}\Xi_{21}^-(K')\right]
\right.\nonumber\\
   &+& \left. 
      2\left[\gamma^{\mu}\Xi_{21}^-(K) \gamma^{\nu}\Xi_{12}^+ (K')
    + \gamma^{\mu}\Xi_{21}^+(K) \gamma^{\nu}\Xi_{12}^-(K')\right]\right\},
\label{Pi99-2} \\ 
\Pi_{99,b}^{\mu\nu}(P) &=&  
\frac{e^2T}{9}\sum_n\int \frac{d^3 {\mathbf k}}{(2\pi)^3}
\mbox{Tr}_{\rm D}  \left\{ 4\left[\gamma^{\mu} G_{3}^+(K) \gamma^{\nu}
G_{3}^+(K')  + \gamma^{\mu} G_{3}^-(K) \gamma^{\nu}G_{3}^-(K')\right]
\right.\nonumber\\
   &+& \left. \gamma^{\mu} G_{4}^+(K) 
\gamma^{\nu}G_{4}^+(K')  + \gamma^{\mu} G_{4}^-(K) \gamma^{\nu}G_{4}^-(K')
\right\}.
\label{Pi99b}
\end{eqnarray}
\end{subequations}
Let us first start with the contribution of the unpaired blue quarks 
$\Pi_{99,b}^{\mu\nu}(P)$. This is proportional to the standard HDL result, 
$\Pi_{99,b}^{\mu\nu}(P) = (10\alpha/9\alpha_s)\Pi_{\rm HDL}^{\mu\nu}(P)$
where $\alpha\equiv e^2/4\pi$ is the fine structure constant and 
$\Pi_{\rm HDL}^{\mu\nu}(P)$ is given in Eq.~(\ref{Pi-HDL}). 
At $p_0=0$ and $p\to 0$, we arrive at
\begin{eqnarray}
\Pi_{99,b}^{\mu\nu}(0) = - \frac{40 \alpha}{9\pi} {\bar \mu}^2
u^{\mu}u^{\nu}.\label{scr99-1}
\end{eqnarray}
To calculate the contribution of the paired quarks ${\tilde \Pi}_{99}^{\mu\nu}(P)$,
we use the same approach as in the previous section. After performing the Matsubara 
frequency summation, we obtain the following representation: 
\begin{eqnarray}
\tilde{\Pi}_{99}^{\mu\nu}(P) &=& \frac{8\pi\alpha}{9}
\int \frac{d^3 {\mathbf k}}{(2\pi)^3}
  \left[
   \left(4 C_{++}^{11} + C_{++}^{22}\right){\cal T}_{++}^{\mu\nu}
  + \left(4 C_{--}^{11} + C_{--}^{22}\right){\cal T}_{--}^{\mu\nu}\right.\nonumber\\
   &+& \left. 
   \left( 4 C_{+-}^{11} + C_{+-}^{22}\right){\cal T}_{+-}^{\mu\nu}
  + \left(4 C_{-+}^{11} + C_{-+}^{22}\right){\cal T}_{-+}^{\mu\nu}\right.\nonumber\\
   &+& \left. 
   2 \left(C_{++}^{12} + C_{++}^{21}\right){\cal U}_{++}^{\mu\nu}
  +2  \left(C_{--}^{12} + C_{--}^{21}\right){\cal U}_{--}^{\mu\nu}\right.\nonumber\\
   &+& \left. 
   2 \left(C_{+-}^{12} + C_{+-}^{21}\right){\cal U}_{+-}^{\mu\nu}
  + 2 \left(C_{-+}^{12} + C_{-+}^{21}\right){\cal U}_{-+}^{\mu\nu} \right].
\label{Pi99-n}
\end{eqnarray} 
By substituting the results in Eqs. (\ref{tu-integral}) and (\ref{c-integral}) into
Eq.~(\ref{Pi99-n}),
we derive the following expression for the polarization tensor $\tilde{\Pi}_{99}^{\mu\nu}$
in the static long-wavelength limit: 
\begin{eqnarray}
\tilde{\Pi}_{99}^{\mu\nu}(0) &\simeq & \frac{16\alpha}{27\pi} \int d k k^2
  \left[3 \left(5C_{++}^{11} - 4C_{+-}^{12} \right)u^{\mu}u^{\nu}
  -\left(5C_{++}^{11}+10C_{+-}^{11}+4C_{+-}^{12}\right)
   \left(g^{\mu\nu}-u^{\mu}u^{\nu}\right) \right] \nonumber\\
&\simeq & -\frac{8\alpha \bar\mu^2}{9\pi} \left[1+\frac{9\delta\mu\, 
\theta(\delta\mu-\Delta)}
{\sqrt{{\delta\mu}^2-\Delta^2}} \right] u^{\mu}u^{\nu} 
- \frac{8\alpha \bar\mu^2}{27\pi} \left[1-\frac{\delta\mu\, 
\theta(\delta\mu-\Delta)}
{\sqrt{{\delta\mu}^2-\Delta^2}} \right] 
\left(g^{\mu\nu}-u^{\mu}u^{\nu}\right).
\label{scr99-2}
\end{eqnarray}
By combining the results in Eqs.~(\ref{scr99-1}) and (\ref{scr99-2}), we obtain the
expression for the photon polarization tensor
\begin{eqnarray}
\Pi_{99}^{\mu\nu}(0) &\simeq &
 -\frac{8\alpha \bar\mu^2}{3\pi} \left[1+\frac{3\delta\mu\, 
\theta(\delta\mu-\Delta)}
{2\sqrt{{\delta\mu}^2-\Delta^2}} \right] u^{\mu}u^{\nu}
- \frac{4\alpha \bar\mu^2}{27\pi} \left[1-\frac{\delta\mu\, 
\theta(\delta\mu-\Delta)}
{\sqrt{{\delta\mu}^2-\Delta^2}} \right] 
\left(g^{\mu\nu}-u^{\mu}u^{\nu}\right).
\label{scr99}
\end{eqnarray}
Now, we calculate the gluon-photon mixed components of the
polarization tensor. The corresponding expression is given by
\begin{subequations}
\begin{eqnarray}
\Pi_{89}^{\mu\nu}(P) &=& \Pi_{98}^{\mu\nu}(P)= 
\frac{1}{2}{\tilde \Pi}_{89}^{\mu\nu}(P)+ 
\frac{1}{2}\Pi_{89,b}^{\mu\nu}(P), \label{Pi89-1} 
\label{Pi89} \\ 
{\tilde \Pi}_{89}^{\mu\nu}(P)
&=& \frac{eg T}{3\sqrt{3}}\sum_n\int \frac{d^3 {\mathbf k}}{(2\pi)^3}
\mbox{Tr}_{\rm D}  \left\{
   2 \left[\gamma^{\mu} G_{1}^+(K) \gamma^{\nu} G_{1}^+(K')  
    + \gamma^{\mu} G_{1}^-(K) \gamma^{\nu}G_{1}^-(K')\right]\right.\nonumber\\
   &-& \left. 
       \gamma^{\mu} G_{2}^+(K) \gamma^{\nu}G_{2}^+(K')  
    - \gamma^{\mu} G_{2}^-(K) \gamma^{\nu}G_{2}^-(K') \right.\nonumber\\
   &+& \left. 
      \gamma^{\mu}\Xi_{12}^-(K) \gamma^{\nu}\Xi_{21}^+(K') 
    + \gamma^{\mu}\Xi_{12}^+(K) \gamma^{\nu}\Xi_{21}^-(K')\right.\nonumber\\
   &-& \left. 
      2 \left[\gamma^{\mu}\Xi_{21}^-(K) \gamma^{\nu}\Xi_{12}^+ (K')
    + \gamma^{\mu}\Xi_{21}^+(K) \gamma^{\nu}\Xi_{12}^-(K')\right]\right\},
\label{Pi89-2} \\ 
\Pi_{89,b}^{\mu\nu}(P) &=&  \frac{egT}{3\sqrt{3}}\sum_n\int \frac{d^3 {\mathbf k}}{(2\pi)^3}
\mbox{Tr}_{\rm D}  \left\{ -2\left[\gamma^{\mu} G_{3}^+(K) \gamma^{\nu}
G_{3}^+(K')  + \gamma^{\mu} G_{3}^-(K) \gamma^{\nu}G_{3}^-(K')\right]
\right.\nonumber\\
   &+& \left. \gamma^{\mu} G_{4}^+(K) 
\gamma^{\nu}G_{4}^+(K')  + \gamma^{\mu} G_{4}^-(K) \gamma^{\nu}G_{4}^-(K')
\right\}.
\label{Pi89b}
\end{eqnarray}
\end{subequations}
The contribution of the unpaired blue quarks is proportional to the HDL expression, 
\begin{equation}
\Pi_{89,b}^{\mu\nu}(P) = -\frac{2}{3}\sqrt{\frac{\alpha}{3\alpha_s}}
\Pi_{\rm HDL}^{\mu\nu}(P),
\end{equation}
with $\Pi_{\rm HDL}^{\mu\nu}(P)$ defined in Eq.~(\ref{Pi-HDL}).
At $p_0=0$ and $p\to 0$, we obtain
\begin{eqnarray}
\Pi_{89,b}^{\mu\nu}(0) 
= \frac{8 \sqrt{\alpha\alpha_s}}{3\sqrt{3}\pi} {\bar \mu}^2
u^{\mu}u^{\nu}.
\label{scr89-1}
\end{eqnarray}
To calculate the contribution of the paired quarks, ${\tilde \Pi}_{89}^{\mu\nu}$,
we first perform the Matsubara frequency summation. The result is
\begin{eqnarray}
\tilde{\Pi}_{89}^{\mu\nu}(P) &=& \frac{4\pi\sqrt{\alpha\alpha_s}}{3\sqrt{3}}
\int \frac{d^3 {\mathbf k}}{(2\pi)^3}
  \left[
   \left(2 C_{++}^{11} - C_{++}^{22}\right){\cal T}_{++}^{\mu\nu}
  + \left(2 C_{--}^{11} - C_{--}^{22}\right){\cal T}_{--}^{\mu\nu}\right.\nonumber\\
   &+& \left. 
   \left( 2 C_{+-}^{11} - C_{+-}^{22}\right){\cal T}_{+-}^{\mu\nu}
  + \left(2 C_{-+}^{11} - C_{-+}^{22}\right){\cal T}_{-+}^{\mu\nu}\right.\nonumber\\
   &+& \left. 
    \left(C_{++}^{12} -2 C_{++}^{21}\right){\cal U}_{++}^{\mu\nu}
  + \left(C_{--}^{12} -2 C_{--}^{21}\right){\cal U}_{--}^{\mu\nu}\right.\nonumber\\
   &+& \left. 
   \left(C_{+-}^{12} -2 C_{+-}^{21}\right){\cal U}_{+-}^{\mu\nu}
  +  \left(C_{-+}^{12} -2 C_{-+}^{21}\right){\cal U}_{-+}^{\mu\nu} \right].
\label{Pi89-n}
\end{eqnarray} 
By substituting the results in Eqs. (\ref{tu-integral}) and (\ref{c-integral}) into
Eq.~(\ref{Pi89-n}),
we get the following expression for the polarization tensor $\tilde{\Pi}_{89}^{\mu\nu}$
in the static long-wavelength limit: 
\begin{eqnarray}
\tilde{\Pi}_{89}^{\mu\nu}(0) &\simeq & 
\frac{8\sqrt{\alpha\alpha_s}}{9\sqrt{3}\pi} 
\int d k k^2
  \left[3 \left(C_{++}^{11} + C_{+-}^{12} \right)u^{\mu}u^{\nu}
  -\left(C_{++}^{11}+2C_{+-}^{11}-C_{+-}^{12}\right)
   \left(g^{\mu\nu}-u^{\mu}u^{\nu}\right) \right] \nonumber\\
&\simeq &  -\frac{8\sqrt{\alpha\alpha_s} \bar\mu^2}{3\sqrt{3}\pi} 
u^{\mu}u^{\nu}
-\frac{8\sqrt{\alpha\alpha_s} \bar\mu^2}{9\sqrt{3}\pi} 
\left[ 1-
\frac{\delta\mu\, \theta(\delta\mu-\Delta)}{\sqrt{{\delta\mu}^2-\Delta^2}} 
\right]
\left( g^{\mu\nu}-u^{\mu}u^{\nu}\right),
\label{scr89-2}
\end{eqnarray}
[c.f. Eq.~(\ref{scr88-1})]. By substituting the results in Eqs.~(\ref{scr89-1}) 
and (\ref{scr89-2}) into Eq.~(\ref{Pi89}), we obtain the expression for the 
mixed components of the polarization tensor
\begin{eqnarray}
\Pi_{89}^{\mu\nu}(0) &\simeq &
-\frac{4\sqrt{\alpha\alpha_s} \bar\mu^2}{9\sqrt{3}\pi} 
\left[ 1-
\frac{\delta\mu\, \theta(\delta\mu-\Delta)}{\sqrt{{\delta\mu}^2-\Delta^2}} 
\right]
\left( g^{\mu\nu}-u^{\mu}u^{\nu}\right).
\label{scr89}
\end{eqnarray}
It is important to emphasize that the explicit result for the
electrical $00$-component of this mixing gluon-photon
polarization tensor is vanishing at $p_0=0$ and $p\to 0$.
This is similar to the ideal 2SC case considered in Ref.~\cite{SWR}.
(Because of this, one should be careful when interpreting the
results for the Debye screening masses in a different basis
of gauge fields \cite{Litim}.) The expressions for the Debye
screening masses read
\begin{subequations}
\begin{eqnarray}
m_{D,88}^2 &=& \frac{4\alpha_s\bar\mu^2}{\pi} ,
\label{m_D_8} \\
m_{D,\gamma\gamma}^2 &=& \frac{8\alpha\bar\mu^2}{3\pi}
\left(1+\frac{3\delta\mu\, \theta(\delta\mu-\Delta)}
{2\sqrt{(\delta\mu)^2-\Delta^2}} \right).
\label{m_D_gamma}
\end{eqnarray}
\label{m_D_8-gamma}
\end{subequations}
In order to obtain the Meissner screening masses, we first derive
all components of the polarization tensor that span the space
of the 8th gluon and the photon. At $p_0=0$ and
$p\to 0$, the corresponding nonzero components, denoted as
$m_{M,AB}^2$, read
\begin{subequations}
\begin{eqnarray}
m_{M,88}^2 &=& \frac{4\alpha_s\bar\mu^2}{9\pi}
\left(1-\frac{\delta\mu\, \theta(\delta\mu-\Delta)}
{\sqrt{(\delta\mu)^2-\Delta^2}} \right),
\label{m_M_88} \\
m_{M,\gamma\gamma}^2 &=& \frac{4\alpha\bar\mu^2}{27\pi}
\left(1-\frac{\delta\mu\, \theta(\delta\mu-\Delta)}
{\sqrt{(\delta\mu)^2-\Delta^2}} \right) ,
\label{m_M_8-gamma} \\
m_{M,8\gamma}^2 &=&
\frac{4\sqrt{\alpha\alpha_s}\bar\mu^2}{9\sqrt{3}\pi}
\left(1-\frac{\delta\mu\, \theta(\delta\mu-\Delta)}
{\sqrt{(\delta\mu)^2-\Delta^2}} \right),
\label{m_M_gamma-gamma}
\end{eqnarray}
\end{subequations}
and $m_{M,\gamma 8}^2 = m_{M,8 \gamma}^2 $. The mixing disappears
in the basis of the fields in Eq.~(\ref{tilde-A-gamma}).
The Meissner screening mass for the $\tilde{8}$th gluon field is
\begin{equation}
m_{M,\tilde{8}\tilde{8}}^2 = \frac{4(3\alpha_s+\alpha)\bar\mu^2}{27\pi}
\left(1-\frac{\delta\mu\, \theta(\delta\mu-\Delta)}
{\sqrt{(\delta\mu)^2-\Delta^2}} \right),
\label{m_M_8}
\end{equation}
and the Meissner screening mass for the medium photon $\tilde{\gamma}$
is vanishing, $m_{M,\tilde{\gamma}\tilde{\gamma}}^2 = 0$. This is consistent with
the absence of the Meissner effect for the
unbroken $\tilde{\rm U}(1)_{\rm em}$.

As is easy to see from Eq.~(\ref{m_M_8}), the medium modified
$\tilde{8}$th gluon has a chromomagnetic instability in the gapless
2SC phase. This is because the Meissner screening mass squared is
{\em negative} when $0<\Delta/\delta\mu<1$. 

\section{Gluons with $A=4,5,6,7$}
\label{gpt44}

After performing the traces over the color, the flavor and the Nambu-Gorkov indices,
the diagonal components of the polarization tensor $\Pi_{AB}^{\mu\nu}(P)$ with $A=B=4,5,6,7$
have the form
\begin{eqnarray}
\Pi_{44}^{\mu\nu}(P) & = & 
\frac{g^2 T}{8} \int \frac{d^3 {\mathbf k}}{(2\pi)^3}{\rm Tr}_{\rm D}
\big[ \gamma^{\mu} G_{3}^+(K) \gamma^{\nu}G_{1}^+(K') 
+ \gamma^{\mu} G_{1}^+(K) \gamma^{\nu} G_{3}^+(K') \nonumber \\
& + & \gamma^{\mu} G_{4}^+(K) \gamma^{\nu}G_{2}^+(K')
+ \gamma^{\mu} G_{2}^+(K) \gamma^{\nu}G_{4}^+(K') \nonumber \\
& + & \gamma^{\mu} G_{3}^-(K) \gamma^{\nu}G_{1}^-(K') 
+ \gamma^{\mu} G_{1}^-(K) \gamma^{\nu}G_{3}^-(K')\nonumber \\
& + & \gamma^{\mu} G_{4}^-(K) \gamma^{\nu}G_{2}^-(K') 
+ \gamma^{\mu} G_{2}^-(K) \gamma^{\nu}G_{4}^-(K') \big].
\end{eqnarray}
[Note that $\Pi_{44}^{\mu\nu}(P)=\Pi_{55}^{\mu\nu}(P)
=\Pi_{66}^{\mu\nu}(P)=\Pi_{77}^{\mu\nu}(P)$.]
Apart from the diagonal elements, there are also nonzero 
off-diagonal elements,
\begin{eqnarray}
\Pi_{45}^{\mu\nu}(P) = - \Pi_{54}^{\mu\nu} (P)
= \Pi_{67}^{\mu\nu}(P) = - \Pi_{76}^{\mu\nu}(P)
= i {\hat \Pi}^{\mu\nu}(P), 
\label{odd-diag}
\end{eqnarray}
with 
\begin{eqnarray}
{\hat \Pi}^{\mu\nu}(P) &=& 
\frac{g^2 T}{8} \int \frac{d^3 {\mathbf k}}{(2\pi)^3}{\rm Tr}_{\rm D}
\big[ \gamma^{\mu} G_{3}^+(K) \gamma^{\nu}G_{1}^+(K') 
- \gamma^{\mu} G_{1}^+(K) \gamma^{\nu} G_{3}^+(K') \nonumber \\
&+& \gamma^{\mu} G_{4}^+(K) \gamma^{\nu}G_{2}^+(K')
- \gamma^{\mu} G_{2}^+(K) \gamma^{\nu}G_{4}^+(K') \nonumber \\
& - & \gamma^{\mu} G_{3}^-(K) \gamma^{\nu}G_{1}^-(K') 
+ \gamma^{\mu} G_{1}^-(K) \gamma^{\nu}G_{3}^-(K')\nonumber \\
& - & \gamma^{\mu} G_{4}^-(K) \gamma^{\nu}G_{2}^-(K') 
+ \gamma^{\mu} G_{2}^-(K) \gamma^{\nu}G_{4}^-(K') \big], 
\end{eqnarray}
The off-diagonal components of the gluon self-energy in Eq.~(\ref{odd-diag})
are nonzero in general. The physical gluon fields in the 2SC/g2SC phase
are the following linear combinations:
$\tilde{A}_{4,5}^{\mu}=(A_{4}^{\mu}\pm i A_{5}^{\mu})/\sqrt{2}$
and  $\tilde{A}_{6,7}^{\mu}=(A_{6}^{\mu}\pm i A_{7}^{\mu})/\sqrt{2}$ 
\cite{R-meissner}. These new fields, 
$\tilde{A}_{4,6}^{\mu}$ and $\tilde{A}_{7,5}^{\mu}$, 
describe two pairs of massive vector particles with well defined 
electomagnetic charges, $\tilde{Q}=\pm 1$. 
The components of the polarization tensor in the new
basis read
\begin{subequations}
\begin{eqnarray}
\tilde{\Pi}_{44}^{\mu\nu}(P) & = & \tilde{\Pi}_{66}^{\mu\nu}(P) = \Pi_{44}^{\mu\nu}(P) + 
{\hat \Pi}^{\mu\nu}(P)  \nonumber \\
& = & 
\frac{g^2 T}{4} \int \frac{d^3 {\mathbf k}}{(2\pi)^3}{\rm Tr}_{\rm D}
 \big[ 
\gamma^{\mu} G_{3}^+(K)
 \gamma^{\nu}G_{1}^+(K')  + \gamma^{\mu} G_{1}^-(K) 
\gamma^{\nu}G_{3}^-(K') \nonumber \\
& & + \gamma^{\mu} G_{4}^+(K) \gamma^{\nu}G_{2}^+(K') + 
\gamma^{\mu} G_{2}^-(K) \gamma^{\nu}G_{4}^-(K') \big], 
\end{eqnarray}
and 
\begin{eqnarray}
\tilde{\Pi}_{55}^{\mu\nu}(P) & = & \tilde{\Pi}_{77}^{\mu\nu}(P) = \Pi_{44}^{\mu\nu} (P)
- {\hat \Pi}^{\mu\nu}(P) \nonumber \\
& = & 
\frac{g^2 T}{4} \int \frac{d^3 {\mathbf k}}{(2\pi)^3}{\rm Tr}_{\rm D}
 \big[ 
\gamma^{\mu} G_{1}^+(K)\gamma^{\nu}G_{3}^+(K')  + \gamma^{\mu} 
G_{3}^-(K) \gamma^{\nu}G_{1}^-(K') \nonumber \\
& & + \gamma^{\mu} G_{2}^+(K) \gamma^{\nu}G_{4}^+(K') + 
\gamma^{\mu} G_{4}^-(K) \gamma^{\nu}G_{2}^-(K') \big].
\end{eqnarray}
\end{subequations}
After Matsubara frequency summation, the polarization tensors 
can be written in the following form:
\begin{subequations}
\begin{eqnarray}
\tilde{\Pi}_{44}^{\mu\nu}(P) & = & \pi\alpha_s
\int \frac{d^3 {\mathbf k}}{(2\pi)^3}
\left[C_{++}^{44} {\cal T}^{\mu\nu}_{++}
+C_{--}^{44} {\cal T}^{\mu\nu}_{--}
+C_{+-}^{44} {\cal T}^{\mu\nu}_{+-}
+C_{-+}^{44} {\cal T}^{\mu\nu}_{-+}\right], 
\label{Pi44-n} \\
\tilde{\Pi}_{55}^{\mu\nu}(P) & = & \pi\alpha_s
\int \frac{d^3 {\mathbf k}}{(2\pi)^3}
\left[C_{++}^{55} {\cal T}^{\mu\nu}_{++}
+C_{--}^{55} {\cal T}^{\mu\nu}_{--}
+C_{+-}^{55} {\cal T}^{\mu\nu}_{+-}
+C_{-+}^{55} {\cal T}^{\mu\nu}_{-+}\right].
\label{Pi55-n}
\end{eqnarray}
\label{Pi44-55}
\end{subequations}
The explicit expressions for the coefficient functions $C_{\pm\pm}^{44,55}$ and 
$C_{\pm\mp}^{44,55}$ at zero and nonzero temperatures are given in Appendix~\ref{Cof-TU44}.
Here we quote only the approximate zero temperature results at $p_0=0$ and $p\to0$. The 
particle-hole contributions are
\begin{eqnarray}
C_{\pm\pm}^{44,55} &\simeq&
-\frac{1}{E_{\Delta,k}^-}
\frac{E_{\Delta,k}^-- E_{k}^-}{E_{\Delta,k}^- +  E_{b,k}^-} 
-\big[ \theta(-E_{bu,k}^-)+\theta(-E_{bd,k}^-) \big] 
\frac{ E_{b,k}^- + E_{k}^-}{(E_{\Delta,k}^-)^2 - ( E_{b,k}^-)^2 }
\nonumber \\
&+& \frac{\theta(-E_{\Delta,k}^-+\delta\mu)}{E_{\Delta,k}^-}
\frac{(E_{\Delta,k}^-)^2 +  E_{k}^-  E_{b,k}^-}
{(E_{\Delta,k}^-)^2 - ( E_{b,k}^-)^2},\label{cccc}
\end{eqnarray}
and the particle-antiparticle contributions are
\begin{eqnarray}
C_{\pm\mp}^{44,55} &\simeq&
-2\left(\frac{|k - \bar\mu | + k - \bar\mu }
{ |k - \bar\mu | (k + \bar\mu_b + |k - \bar\mu |)} -\frac{1}{k}\right) ,
\label{C_44,55}
\end{eqnarray}
where $E_{b,k}^\pm\equiv E_{k}\pm \bar\mu_b$ and 
$\bar\mu_b=\bar\mu-\mu_8$. The vacuum contribution was subtracted 
in Eq.~(\ref{C_44,55}) in order to remove ultraviolet divergences 
in the polarization tensor.

To proceed, we need to calculate the following two types of momentum 
integrals (for the details of the calculation see Appendix~\ref{Int-44}):
\begin{subequations}
\begin{eqnarray}
\int d k k^2 C_{\pm\pm}^{44,55} &\simeq& 
-2\bar\mu^2\left[1-\frac{\Delta^2}{4\mu_8^2}\ln
\frac{\left(\Delta^2+\mu_8^2\right)^2-4\left(\mu_8\delta\mu\right)^2}{\Delta^4}
-\frac{\Delta^2}{4\mu_8^2}\ln
\frac{\Delta^4-\mu_8^2\left(\delta\mu-\sqrt{\delta\mu^2-\Delta^2}\right)^2}
{\Delta^4-\mu_8^2\left(\delta\mu+\sqrt{\delta\mu^2-\Delta^2}\right)^2}
\theta\left(\delta\mu-\Delta\right)\right]
\nonumber\\
&\approx& -2\bar\mu^2\left[
\frac{\Delta^2+2(\delta\mu)^2}{2\Delta^2}
-\frac{\delta\mu\sqrt{\delta\mu^2-\Delta^2}}{\Delta^2}
\theta\left(\delta\mu-\Delta\right)
\right],\quad\mbox{for}\quad \mu_8\to 0,
\label{c4pp-integral}
\end{eqnarray}
and 
\begin{eqnarray}
\int d k k^2 C_{\pm\mp}^{44,55} &\simeq& \bar\mu^2.
\label{c4pm-integral}
\end{eqnarray}
\label{c4-all}
\end{subequations}
By substituting the results in Eqs.~(\ref{tu-integral}) and (\ref{c4-all}) 
into Eq.~(\ref{Pi44-55}),
we derive expressions for the polarization tensors $\tilde{\Pi}_{44,55}^{\mu\nu}$
in the static long-wavelength limit: 
\begin{eqnarray}
&&\tilde{\Pi}_{44}^{\mu\nu}(0) \simeq  \tilde{\Pi}_{55}^{\mu\nu}(0) 
=\frac{2\alpha_s}{3\pi} \int d k k^2\left[3 C_{++}^{44}u^{\mu}u^{\nu}
-\left(C_{++}^{44}+2C_{+-}^{44}\right)\left(g^{\mu\nu}-u^{\mu}u^{\nu}\right)\right]
\nonumber\\
&=& - \frac{4\alpha_s\bar\mu^2}{\pi} \left[
1-\frac{\Delta^2}{4\mu_8^2}\ln
\frac{\left(\Delta^2+\mu_8^2\right)^2-4\left(\mu_8\delta\mu\right)^2}{\Delta^4}
-\frac{\Delta^2}{4\mu_8^2}\ln
\frac{\Delta^4-\mu_8^2\left(\delta\mu-\sqrt{\delta\mu^2-\Delta^2}\right)^2}
{\Delta^4-\mu_8^2\left(\delta\mu+\sqrt{\delta\mu^2-\Delta^2}\right)^2}
\theta\left(\delta\mu-\Delta\right)
\right] u^{\mu}u^{\nu}\nonumber\\
&-&\frac{\alpha_s\bar\mu^2}{3\pi}
\frac{\Delta^2}{\mu_8^2} \left[\ln
\frac{\left(\Delta^2+\mu_8^2\right)^2-4\left(\mu_8\delta\mu\right)^2}{\Delta^4}
+\ln
\frac{\Delta^4-\mu_8^2\left(\delta\mu-\sqrt{\delta\mu^2-\Delta^2}\right)^2}
{\Delta^4-\mu_8^2\left(\delta\mu+\sqrt{\delta\mu^2-\Delta^2}\right)^2}
\theta\left(\delta\mu-\Delta\right)
\right]
\left(g^{\mu\nu}-u^{\mu}u^{\nu}\right).
\end{eqnarray}
In the neutral 2SC/g2SC phase of quark matter, the value of the color chemical 
potential $\mu_8$ is small \cite{HS}. Therefore, a good approximation for the above 
polarization tensors is obtained by taking the limit $\mu_8\to 0$,
\begin{eqnarray}
\tilde{\Pi}_{44}^{\mu\nu}(0) =  \tilde{\Pi}_{55}^{\mu\nu}(0) 
&\approx& - \frac{2\alpha_s\bar\mu^2}{\pi}\left[
\frac{\Delta^2+2(\delta\mu)^2}{\Delta^2}
-2\frac{\delta\mu\sqrt{\delta\mu^2-\Delta^2}}{\Delta^2}
\theta\left(\delta\mu-\Delta\right)
\right]u^{\mu}u^{\nu} \nonumber\\
&&-\frac{2\alpha_s\bar\mu^2}{3\pi}\left[
\frac{\Delta^2-2(\delta\mu)^2}{\Delta^2}
+2\frac{\delta\mu\sqrt{\delta\mu^2-\Delta^2}}{\Delta^2}
\theta\left(\delta\mu-\Delta\right)
\right]\left(g^{\mu\nu}-u^{\mu}u^{\nu}\right).
\end{eqnarray}
Therefore, the corresponding Debye and Meissner screening masses are
\begin{subequations}
\begin{eqnarray}
m_{D,44}^2 &=& m_{D,55}^2
= \frac{2\alpha_s\bar\mu^2}{\pi}\left[
\frac{\Delta^2+2(\delta\mu)^2}{\Delta^2}
-2\frac{\delta\mu\sqrt{\delta\mu^2-\Delta^2}}{\Delta^2}
\theta\left(\delta\mu-\Delta\right)
\right],
\label{Debye-44} \\
m_{M,44}^2 &=& m_{M,55}^2 =  \frac{2\alpha_s\bar\mu^2}{3\pi}\left[
\frac{\Delta^2-2(\delta\mu)^2}{\Delta^2}
+2\frac{\delta\mu\sqrt{\delta\mu^2-\Delta^2}}{\Delta^2}
\theta\left(\delta\mu-\Delta\right)
\right].
\label{Meissner-44} 
\end{eqnarray}
\end{subequations}
It is clear from Eq.~(\ref{Meissner-44}) that the corresponding four 
gluons have chromomagnetic (plasma type) instabilities in the gapless
2SC phase ($0<\Delta/\delta\mu<1$), as well as in the gapped 2SC 
phase when $1<\Delta/\delta\mu<\sqrt{2}$. This is indicated by a {\em negative}
value of the Meissner screening mass squared.

\section{Discussion}
\label{discussion}

The results of our calculation for the screening masses in neutral two-flavor 
quark matter are summarized in Table~\ref{DM-masses}. There we list the squared 
values of the Debye and Meissner screening masses in units of
$m_g^2 \equiv {4\alpha_s\bar\mu^2} / {3\pi}$. 

\begin{table}
\begin{tabular}{|l|c|c|}
\hline
\hline
 & $m_{D,x}^2/m_g^2$ & $m_{M,x}^2/m_g^2$ \\[1mm]
\hline
\hline ~ & & \\[-3mm]
$x=(11),(22),(33)$ &$\frac{3\delta\mu}
{\sqrt{(\delta\mu)^2-\Delta^2}} \theta(\delta\mu-\Delta) $ & $0$\\[3mm]
\hline ~ & & \\[-3mm]
$x=(44),(55),(66),(77)$ & $3 \frac{\Delta^2+2\delta\mu^2}{2\Delta^2}
-3\frac{\delta\mu\sqrt{\delta\mu^2-\Delta^2}}{\Delta^2} 
\theta(\delta\mu-\Delta)$& $\frac{\Delta^2-2\delta\mu^2}{2\Delta^2}
+\frac{\delta\mu\sqrt{\delta\mu^2-\Delta^2}}{\Delta^2}
\theta(\delta\mu-\Delta) $\\[3mm]
\hline ~ & & \\[-3mm]
$x=(88)$ & $3$ & $\frac{1}{3}
\left(1-\frac{\delta\mu\, \theta(\delta\mu-\Delta)}
{\sqrt{(\delta\mu)^2-\Delta^2}} \right)$ \\[3mm]
\hline ~ & & \\[-3mm]
$x=(8\gamma),(\gamma 8)$ & $0$ & $\frac{\sqrt{\alpha}}{3\sqrt{3\alpha_s}}
\left(1-\frac{\delta\mu\, \theta(\delta\mu-\Delta)}
{\sqrt{(\delta\mu)^2-\Delta^2}} \right)$\\[3mm]
\hline ~ & & \\[-3mm]
$x=(\gamma\gamma) $ & $\frac{2\alpha}{\alpha_s}
\left(1+\frac{3\delta\mu\, \theta(\delta\mu-\Delta)}
{2\sqrt{(\delta\mu)^2-\Delta^2}} \right)$ & $\frac{\alpha}{9\alpha_s}
\left(1-\frac{\delta\mu\, \theta(\delta\mu-\Delta)}
{\sqrt{(\delta\mu)^2-\Delta^2}} \right)$\\[3mm]
\hline\hline
~ & & \\[-3mm]
$x=(\tilde{8}\tilde{8})$ 
& $\frac{9\alpha_s}{3\alpha_s+\alpha}\left[
1+\frac{2\alpha^2}{9\alpha_s^2}\left(1+\frac{3\delta\mu\, \theta(\delta\mu-\Delta)}
{2\sqrt{(\delta\mu)^2-\Delta^2}} \right)\right]$ 
& $\frac{3\alpha_s+\alpha}{9\alpha_s}
\left(1-\frac{\delta\mu\, \theta(\delta\mu-\Delta)}
{\sqrt{(\delta\mu)^2-\Delta^2}} \right)$ \\[3mm]
\hline ~ & & \\[-3mm]
$x=(\tilde{8}\tilde{\gamma}),(\tilde{\gamma} \tilde{8})$ 
& $-\frac{3\sqrt{3\alpha\alpha_s}}{3\alpha_s+\alpha}\left[
1-\frac{2\alpha}{3\alpha_s}\left(1+\frac{3\delta\mu\, \theta(\delta\mu-\Delta)}
{2\sqrt{(\delta\mu)^2-\Delta^2}} \right)\right]$ 
& $0$\\[3mm]
\hline ~ & & \\[-3mm]
$x=(\tilde{\gamma}\tilde{\gamma}) $ 
& $\frac{9\alpha}{3\alpha_s+\alpha}\left(
1+\frac{\delta\mu\, \theta(\delta\mu-\Delta)}
{\sqrt{(\delta\mu)^2-\Delta^2}} \right)$ 
& $0$\\[3mm]
\hline
\hline
\end{tabular}
\caption{\label{DM-masses} The Debye and Meissner screening masses for gauge bosons
in neutral two-flavor quark matter. By definition, 
$m_g^2 \equiv 4\alpha_s\bar\mu^2/3\pi$.}
\end{table}

\subsection{Debye screening masses}
\label{dsm}

\begin{figure}
\begin{minipage}[t]{0.442\textwidth}
\includegraphics[width=8cm]{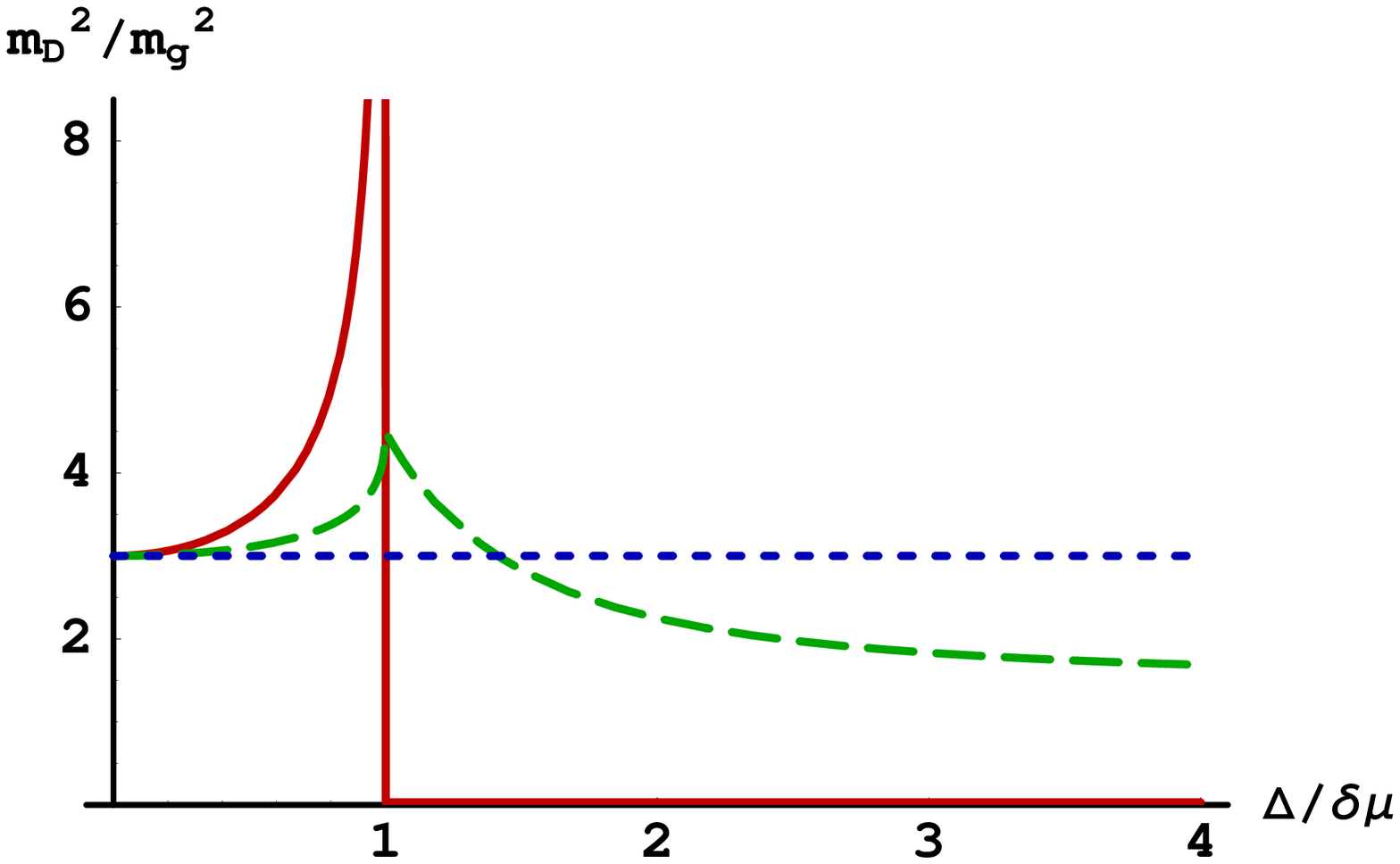}
\caption{\label{fig-D-masses}
Squared values of the Debye screening masses as functions of the 
dimensionless ratio $\Delta/\delta\mu$ for the gluons with 
$A=1,2,3$ (solid line), for the gluons with $A=4,5,6,7$ 
(long-dashed line), and for the $\tilde{8}$th gluon (short-dashed 
line). By definition, $m_g^2 \equiv 4\alpha_s\bar\mu^2/3\pi$.}
\end{minipage}
\hspace{0.1\textwidth}
\begin{minipage}[t]{0.442\textwidth}
\includegraphics[width=8cm]{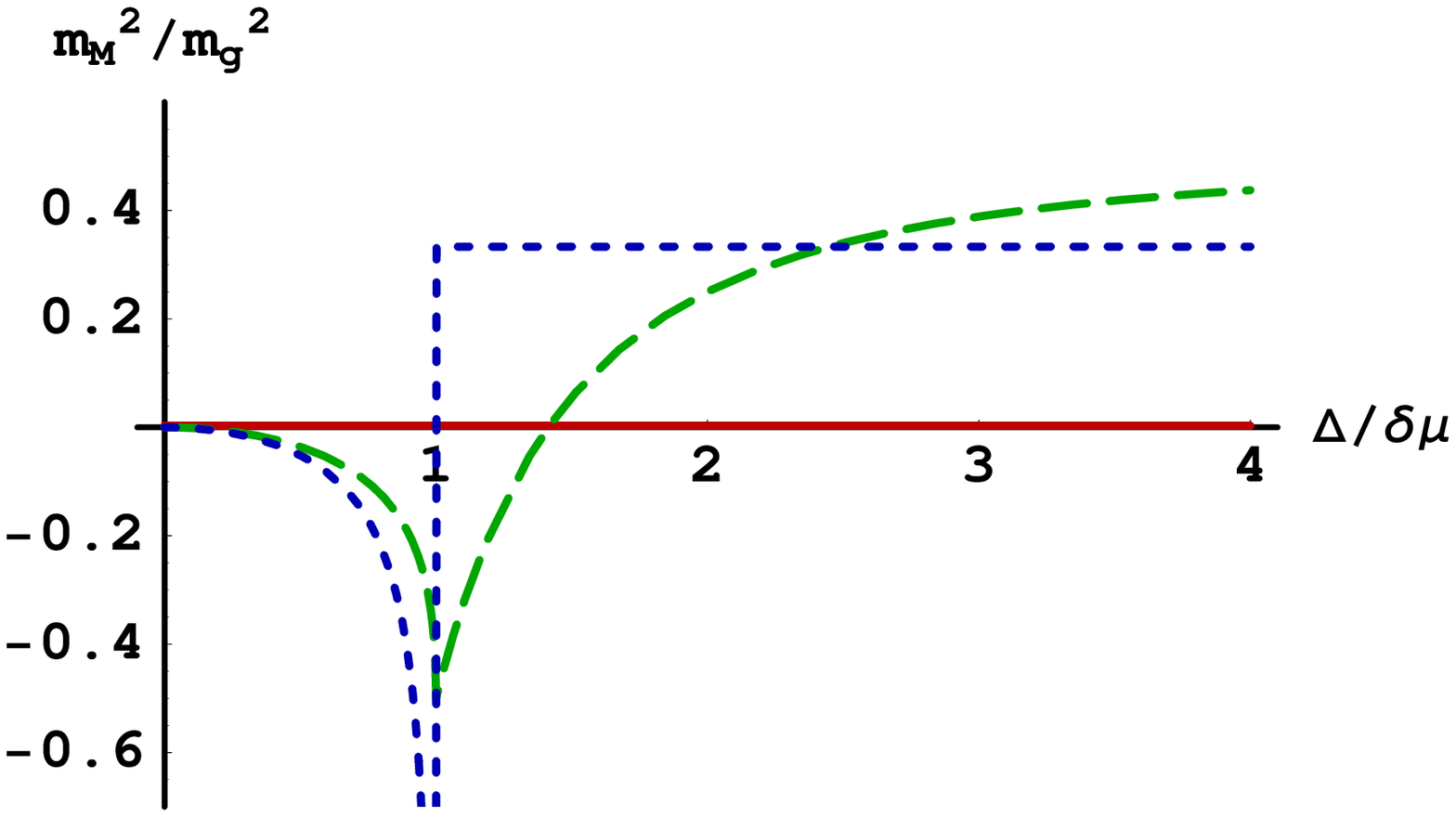}
\caption{\label{fig-M-masses}
Squared values of the Meissner screening masses as functions of the 
dimensionless ratio $\Delta/\delta\mu$ for the gluons with 
$A=1,2,3$ (solid line), for the gluons with $A=4,5,6,7$ 
(long-dashed line), and for the $\tilde{8}$th gluon (short-dashed 
line). By definition, $m_g^2 \equiv 4\alpha_s\bar\mu^2/3\pi$.}
\end{minipage}
\end{figure}

Let us first discuss the results for the Debye screening masses.
The values of $m_{D,x}^2/m_g^2$ with $x=(11),(44),(88)$ as functions 
of the dimensionless ratio $\Delta/\delta\mu$ are shown graphically 
in Fig.~\ref{fig-D-masses} (we do not show the Debye screening mass 
for the photon whose value is suppressed by the fine structure 
constant). The vanishing value of the ratio $\Delta/\delta\mu$ 
corresponds to the normal phase of quark matter. In this limit, 
as is easy to check, our results for the Debye screening masses 
coincide with the known results in Refs.~\cite{Vija,Manuel}. Also, 
in the other limiting case, $\Delta/\delta\mu=\infty$, our results 
coincide with those in the ideal 2SC phase \cite{R-meissner,SWR}. 
As before, by the ideal 2SC phase of quark matter, we mean the 2SC 
phase without any mismatch between the Fermi momenta of the up and 
down quarks, i.e., $\delta\mu=0$.

We see that the value of the Debye screening mass for the gluons of 
the unbroken $SU(2)_c$ gauge subgroup is vanishing in the 2SC phase 
($\Delta/\delta\mu >1$). This result can be understood in the same 
way as in the ideal 2SC case \cite{R-meissner}. It reflects the fact 
that there are no gapless quasiparticles charged with respect to the 
$SU(2)_c$ subgroup in the ground state. In the g2SC phase, on the 
other hand, there are gapless charged quasiparticles around the two 
``effective'' Fermi momenta 
$p_F^{\rm eff} = \mu^{\pm}\equiv \bar\mu\pm\sqrt{(\delta\mu)^2-\Delta^2}$
\cite{SH,HS}. The densities of states at the corresponding Fermi 
surfaces are \cite{HS}
\begin{eqnarray}
\left. \frac{dN}{dE}\right|_{p=\mu^{\pm}}
& = &\frac{2\delta\mu (\mu^{\pm})^2}
{\sqrt{(\delta\mu)^2-\Delta^2}}.
\end{eqnarray}
As one can check from Table~\ref{DM-masses}, the squared value of the Debye 
screening mass $m_{D,11}^2$ is proportional to the sum of these densities of 
the gapless states (up to higher order corrections). Turning these arguments 
around, one could obtain the result for the Debye screening mass without 
calculating the polarization tensor. Indeed, one can write \cite{kemer}
\begin{eqnarray}
m_{D,11}^2 \sim \left. \frac{dN}{dE}\right|_{p=\mu^{-}} 
+ \left. \frac{dN}{dE}\right|_{p=\mu^{+}},
\end{eqnarray}
and determine the overall coefficient $\alpha_s/\pi$ by matching the 
expression on the right hand side with the known expression in the normal 
phase ($\Delta/\delta\mu=0$). 

It is interesting to consider $m_{D,11}^2$ in the limit when the gap 
$\Delta$ approaches $\delta\mu$ from the side of the gapless phase. 
As one can see, the formal value of $m_{D,11}^2$ goes to infinity as 
$\alpha_s \bar\mu^2/\sqrt{1-(\Delta/\delta\mu)^2}$ when $\Delta\to 
\delta\mu-0$. This is the consequence of having a quadratic dispersion 
relation for gapless quasiparticles $E_{\Delta^{-}}^{-}\simeq 
(p-\bar\mu)^2/2\Delta$ when $\Delta\to\delta\mu$. However, the 
infinitely large value of the Debye mass has little physical meaning.
This is because the distance scales, at which the corresponding 
screening is set in, also become infinite in the limit $\Delta\to\delta\mu$.

From Table~\ref{DM-masses}, we see that there is no mixing between the
Debye screening masses of the 8th gluon and of the vacuum photon. It is 
known, however, that the physical modes of the corresponding gauge bosons 
in the 2SC/g2SC phase are the linear combinations given in 
Eq.~(\ref{tilde-A-gamma}) which are different from the vacuum fields. 
Moreover, one of the linear combinations, see Eq.~(\ref{tilde-gamma}), 
describes the medium modified photon of the unbroken electromagnetism. 
One might think that the absence of mixing between the electric screening 
masses is in conflict with the gauge invariance of the 2SC/g2SC ground 
state with respect to $\tilde{\rm U}(1)_{\rm em}$. However, the two 
propagating modes of the low-energy photon $\tilde{\gamma}$ of 
$\tilde{\rm U}(1)_{\rm em}$ have transverse polarizations and, 
therefore, should come from the magnetic sector. The third, electrical 
mode of $\tilde{\gamma}$ is not massless. This latter decouples
from the low-energy theory and its presence is irrelevant for the
gauge invariance with respect to $\tilde{\rm U}(1)_{\rm em}$.

\subsection{Meissner screening masses}
\label{msm}

Now we discuss the results for the Meissner screening masses. The values 
of $m_{M,x}^2/m_g^2$ with $x=(11),(44),(88)$ as functions of the 
dimensionless ratio $\Delta/\delta\mu$ are shown graphically in 
Fig.~\ref{fig-M-masses} (the screening mass for the photon, suppressed 
by the fine structure constant, is not shown). As it should be, the 
results in the two limiting cases, i.e., $\Delta/\delta\mu=0$ and 
$\Delta/\delta\mu=\infty$, coincide with the results in the normal 
phase (i.e., no Meissner effect) and with the results in the ideal 
2SC phase \cite{R-meissner,SWR}, respectively.

In agreement with the group theoretical arguments, the Meissner masses 
of the three gluons of the unbroken $SU(2)_c$ subgroup are vanishing 
in both the gapless and the gapped 2SC phases, see the solid line in 
Fig.~\ref{fig-M-masses}. The vanishing Meissner masses come as a result 
of the exact cancellation between the diamagnetic particle-antiparticle 
and the paramagnetic particle-hole contributions. 

The results in Table~\ref{DM-masses} show that the Meissner screening 
masses in the subspace of the 8th gluon and the vacuum photon have a
nontrivial mixing. As we saw in Sec.~\ref{gpt88}, the mixing disappears 
after switching to the description in terms 
of the physical modes, defined in Eq.~(\ref{tilde-A-gamma}). Moreover, 
the Meissner screening mass of the medium photon is vanishing. This is
consistent with the absence of the Meissner effect for the unbroken 
electromagnetism.

The most interesting results of our calculation are the expressions 
for the Meissner screening masses of the five gluons that correspond 
to the broken generators of the $SU(3)_c$ group. We find that the 
squared values of these masses are {\em negative} (i.e., the masses 
themselves are imaginary) in some regions of parameters. In particular, 
we obtain $m_{M,x}^{2}<0$ for $x=(44),(55),(66),(77)$ when 
$0<\Delta/\delta\mu<\sqrt{2}$, and $m_{M,\tilde{8}\tilde{8}}^{2}<0$ when 
$0<\Delta/\delta\mu<1$. The standard interpretation of such a result 
is the existence of a chromomagnetic (plasma type) instability in the 
corresponding phases of matter \cite{pi}. 

While the instability connected with the $\tilde{8}$th gluon appears 
only in the g2SC phase ($0<\Delta/\delta\mu<1$), the instability 
connected with the other 4 gluons also develops in the {\em gapped} 
2SC phase when $1<\Delta/\delta\mu<\sqrt{2}$. This suggests that the 
gapless superconductivity itself is not the reason for the 
instability. It is interesting to note, however, that the critical 
value of the mismatch, above which the instability starts to develop, is 
given by $\delta\mu_c=\Delta/\sqrt{2}$. This happens to coincide with the 
value of the mismatch at which the 2SC phase becomes metastable when the 
neutrality condition is not enforced in quark matter (i.e., when 
the value of $\delta\mu$ is treated as a free parameter and the Coulomb 
effects are ignored). Here, on the other hand, we consider neutral quark 
matter in the same way as in Refs.~\cite{SH,HS}.

It is natural to suggest that the instability, indicated by negative 
values of the Meissner screening masses squared, may result in some 
type of a gluon condensation. Indeed, in terms of the gluon effective 
potential, the appearance of a negative gluon mass squared could be viewed 
as considering a false vacuum that corresponds to a local maximum of 
the potential at $\langle A_\mu^a \rangle= 0$. In this case, the true 
vacuum should be given by the global minimum of the gluon potential.
It would be natural then if the minimum corresponds to another (nonzero) 
expectation value of the gluon field, i.e., $\langle A_\mu^a 
\rangle\neq 0$ for $a\in (4,5,6,7)$, or for $a=8$.

Note that we do not exclude the possibility that the new stable 
ground state has a condensate that breaks the rotational symmetry 
of the system. In fact, this might be the most natural outcome of 
the gluon condensation if its mechanism is the same, e.g., as in 
Ref.~\cite{GMS,footnote}. The possibility of breaking the rotational 
symmetry is also suggested by the fact the it is the magnetic 
components of gluons $A_i^a$ rather than the electric gluons $A_0^a$ 
that drive the instability. The expectation that the rotational
symmetry is broken in the ground state may also hint at the possibility 
of a state with deformed quark Fermi surfaces which was proposed in 
Ref.~\cite{MutherSed}.

It may happen that not only the rotational symmetry but also the 
translational symmetry is broken in the true ground state of neutral 
two-flavor quark matter. This would be the case when the diquark 
condensate is inhomogeneous like, for example, in the crystalline
phase \cite{LOFF}, or like in the Abrikosov's 
vortex lattice phase \cite{vortice}. Regarding the crystalline
phase (also known as the Larkin-Ovchinnikov-Fulde-Ferrell phase
\cite{loff-orig}) in two-flavor quark matter, it might be appropriate 
to mention that this phase is claimed to appear precisely when
$\Delta/\delta\mu < \sqrt{2}$. It is not clear, however, how a first
order transition from the 2SC phase to the crystalline phase at
$\Delta/\delta\mu =\sqrt{2}$ can be triggered by vanishingly small
tachyonic masses of gluons. 

We would like to emphasize here that the gluon type instability, 
indicated by negative values of the Meissner screening masses squared,
has nothing to do (at least, directly) with the so-called Sarma 
instability \cite{Sarma}. It was shown for the 
first time in Ref.~\cite{SH}, and then confirmed 
in Refs.~\cite{GLW,var-appr,HS,LZ,gCFL,gCFL-long,dSC,RSR} that the Sarma 
instability in the effective potential for the order parameter 
is removed when the neutrality condition is imposed, or when the 
pairing interaction has a specific momentum structure and the
ratio of the densities of the pairing fermions is kept fixed \cite{FGLW}. 
One may still argue that, despite the absence of the Sarma 
instability in the effective potential, there exists another 
type of an instability, discussed in Ref.~\cite{WuYip}. There it 
was suggested that paramagnetic currents should be spontaneously 
induced in the gapless phase. In fact, we think that the instability 
connected with the $\tilde{8}$th gluon in the g2SC phase could indeed 
be of this type.

One may find some similarity between the chromomagnetic instability 
found in this paper and the instabilities that have been discussed in 
Refs.~\cite{Mrow1,RS,Birse,Arnold,Mrow2} in application to a state 
of matter produced in heavy ion collisions. It should be emphasized, 
however, that the quark distribution functions are completely 
isotropic in momentum space in the case of dense quark matter 
studied in this paper which is in contrast to a typical situation in 
Ref.~\cite{Mrow1,RS,Birse,Arnold,Mrow2}. This fact suggests that the 
mechanism of the instability seen here is more subtle. 

In passing, we would like to mention that there might exist some analogy 
between the instability found in this paper and the so-called paramagnetic
superconductivity discussed in condensed matter physics \cite{PME}. 
If these phenomena are related indeed, it would be natural to expect 
that the instability of neutral dense quark matter is resolved through 
a spontaneous chromo-magnetization. Again, the breakdown of the 
rotational symmetry in this case would seem inevitable.

\section{Conclusion}
\label{conclusion}

In this paper, we calculated the Debye and Meissner screening masses 
for the gluons and the photon in the case of neutral, $\beta$-equilibrated 
two-flavor dense quark matter. A general form of our result allows to 
use it in the gapped phase as well as in the gapless color superconducting 
phase by a simple change of the magnitude of the diquark pairing. This latter 
determines the ratio of the gap parameter and the mismatch of the 
quark Fermi momenta, i.e., $\Delta/\delta\mu$. The qualitative dependence 
of this ratio on the diquark coupling constant in the case of neutral 
quark matter was studied in Refs.~\cite{SH,HS}. It was shown that the 
ground state corresponds to the normal phase at weak coupling, i.e., 
$\Delta/\delta\mu=0$. At some intermediate values of the diquark coupling,
the ground state is the g2SC phase. The ratio $\Delta/\delta\mu$ is
less than 1 in such a gapless phase. Finally, at large coupling, the 
ground state is the 2SC phase, and the ratio $\Delta/\delta\mu$ is 
larger than 1. One could also check that this ratio $\Delta/\delta\mu$ 
increases monotonically with increasing the coupling. 

The results for the Debye and the Meissner screening masses in this paper 
give a natural interpolation between the known values in the normal phase 
\cite{Vija,Manuel} and in the ideal 2SC
phase \cite{R-meissner,SWR}. The most important result of our calculation 
is that the expressions for the Meissner screening masses of the five 
gluons, corresponding to the five broken generators of the $SU(3)_c$ group,
are {\em imaginary}.  This is interpreted as an indication of a chromomagnetic 
instability in the corresponding phases of quark matter. It remains
to be clarified the consequences of such an instability, and the nature
of the true ground state in neutral two-flavor quark matter.

In the future, it would be very interesting to investigate whether
a chromomagnetic instability also develops in the gCFL phase 
\cite{gCFL,gCFL-long,RSR}, where the low-energy quasiparticle 
spectrum resembles the spectrum in the g2SC phase.

\begin{acknowledgments}
The authors thank M.~Buballa, M.~Forbes, T.~Hatsuda, C.~Kouvaris, 
J.~Lenaghan, M.S.~Li, V.~Miransky, S.~Mr\'owczy\'nski, R.~Pisarski, 
K.~Rajagopal, A.~Rebhan, S.~Reddy, P.~Reuter, D.~Rischke, P.~Romatschke, 
T.~Sch\"afer, A.~Schmitt, D.~Son, M.~Strickland, M.~Tachibana, 
D.~N.~Voskresensky, and Q.~Wang for interesting
discussions. I.A.S. is grateful to the INT at the University of
Washington in Seattle for its hospitality. The work of M.H. was
supported by the Alexander von Humboldt-Foundation, and the NSFC 
under Grant Nos. 10105005, 10135030. The work of
I.A.S. was supported by Gesellschaft f\"{u}r Schwerionenforschung
(GSI) and by Bundesministerium f\"{u}r Bildung und Forschung (BMBF).
\end{acknowledgments}

\appendix

\section{Calculation of Dirac traces}
\label{TU-matrix}

Here we calculate the traces defined in Eq.~(\ref{D-traces}).
By neglecting corrections of order $p^2/\mu^2$, we derive the 
following results:
\begin{eqnarray}
{\cal T}^{\mu\nu}_{\pm\pm}
=\mbox{tr}\left[\gamma^{\mu} \gamma^0 \Lambda^{(\pm)}_{k}
\gamma^{\nu} \gamma^0 \Lambda^{(\pm)}_{k'}\right] 
&\simeq &
2\left(u^{\mu}u^{\nu}
\mp \frac{u^{\mu} {\mathbf k}^{\nu}+{\mathbf k}^{\mu}u^{\nu}}{k}
+\frac{{\mathbf k}^{\mu}{\mathbf k}^{\nu}}{k^{2}}\right),
\label{tr2++} \\
{\cal T}^{\mu\nu}_{\pm\mp}
=\mbox{tr}\left[\gamma^{\mu} \gamma^0 \Lambda^{(\pm)}_{k}
\gamma^{\nu} \gamma^0 \Lambda^{(\mp)}_{k'}\right]  
&\simeq &
-2\left(g^{\mu\nu}-u^{\mu}u^{\nu}
+\frac{{\mathbf k}^{\mu}{\mathbf k}^{\nu}}{k^{2}}
\right) ,
\label{tr2+-}\\
{\cal U}^{\mu\nu}_{\pm\pm}
=\mbox{tr}\left[\gamma^{\mu} \gamma^{5}\Lambda^{(\pm)}_{k}
\gamma^{\nu} \gamma^{5} \Lambda^{(\pm)}_{k'}\right] 
&\simeq&
- 2\left( g^{\mu\nu} - u^{\mu}u^{\nu}
+ \frac{{\mathbf k}^{\mu}{\mathbf k}^{\nu}}{k^{2}}
\right),
\label{tr1++}\\
{\cal U}^{\mu\nu}_{\pm\mp}
=\mbox{tr}\left[\gamma^{\mu} \gamma^{5} \Lambda^{(\pm)}_{k}
\gamma^{\nu} \gamma^{5} \Lambda^{(\mp)}_{k'}\right]
&\simeq &
- 2 \left( u^{\mu}u^{\nu}
-\frac{{\mathbf k}^{\mu}{\mathbf k}^{\nu}}{k^{2}}
\pm \frac{u^{\mu} {\mathbf k}^{\nu}-{\mathbf k}^{\mu}u^{\nu}}{k}
\right),
\label{tr1+-}
\end{eqnarray}
where ${\mathbf k'}\equiv {\mathbf k}-{\mathbf p}$ and $u^\mu=(1,0,0,0)$.
In calculation of the polarization tensor, we also encounter the following
type angular integrals:
\begin{eqnarray}
\int\frac{d\phi}{2\pi} \frac{{\mathbf k}^{\mu}}{k} 
f(k,p,\xi)
&=& \xi \frac{{\mathbf p}^{\mu}}{p}f(k,p,\xi),
\label{phi-int1}\\
\int\frac{d\phi}{2\pi} \frac{{\mathbf k}^{\mu}{\mathbf k}^{\nu}}
{k^2} f(k,p,\xi)
&=& -\frac{1}{2}f(k,p,\xi)\left[
(1-\xi^2)(g^{\mu\nu}-u^\mu u^\nu) 
+(1-3\xi^2)\frac{{\mathbf p}^{\mu}{\mathbf p}^{\nu}}{p^2}\right],
\label{phi-int2}
\end{eqnarray}
where $\xi$ is the cosine of the angle between the vectors ${\mathbf k}$ 
and ${\mathbf p}$. These relations are used in Eq.~(\ref{traces-xi}).

\section{Coefficient functions used in calculation of polarization tensor}

\subsection{The coefficient functions 
for $\Pi^{AB}_{\mu\nu}$ with $A,B=1,2,3,8,9$}
\label{Cof-TU11}

As we saw in the main part of the paper, see Eqs. (\ref{Pi11-n}), 
(\ref{Pi99-n}), and (\ref{Pi89-n}), all components of the polarization 
tensor $\Pi^{AB}_{\mu\nu}$ with $A,B=1,2,3,8,9$ were given in terms of 
the following coefficient functions (here and below, 
$\alpha=1$ for $C_{e_1e_2}^{11}$ and $C_{e_1e_2}^{12}$, while 
$\alpha=-1$ for $C_{e_1e_2}^{22}$ and $C_{e_1e_2}^{21}$, 
and $e_1,e_2=\pm$):
\begin{eqnarray}
C_{++}^{11,22}=\sum_{\lambda=\pm}&\lambda &
\left[ {\tilde f}(\lambda E_{\Delta,k}^+ 
+ \alpha\delta\mu) \frac{\lambda E_{\Delta,k}^+ 
-  E_k^+}{2 E_{\Delta,k}^+} 
\frac{-p_0 + \lambda E_{\Delta,k}^+ 
- E_{k'}^+}{(-p_0+\lambda E_{\Delta,k}^+)^2 
- (E^+_{\Delta,k'})^2} \right.\nonumber \\
&+& {\tilde f}(\lambda E_{\Delta,k'}^+
+\alpha\delta\mu) \frac{\lambda E_{\Delta,k'}^+ 
- E_{k'}^+}{2 E_{\Delta,k'}^+} 
\frac{p_0 + \lambda E_{\Delta,k'}^+ 
- E_{k}^+}{(p_0+\lambda E_{\Delta,k'}^+)^2 
- (E^+_{\Delta,k})^2} \nonumber \\
& + & {\tilde f}(\lambda E_{\Delta,k}^-
- \alpha\delta\mu) \frac{\lambda E_{\Delta,k}^- 
- E_k^-}{2 E_{\Delta,k}^-} \frac{-p_0 +\lambda E_{\Delta,k}^- 
- E_{k'}^-}{(-p_0+\lambda E_{\Delta,k}^-)^2 
- (E^-_{\Delta,k'})^2} \nonumber \\
& + & \left.
{\tilde f}(\lambda E_{\Delta,k'}^-
- \alpha\delta\mu) \frac{\lambda E_{\Delta,k'}^- 
- E_{k'}^-}{2 E_{\Delta,k'}^-} \frac{p_0 + \lambda E_{\Delta,k'}^- 
- E_{k}^-}{(p_0+ \lambda E_{\Delta,k'}^-)^2 
- (E^-_{\Delta,k})^2}\right],
\end{eqnarray}

\begin{eqnarray}
C_{--}^{11,22}=\sum_{\lambda=\pm}&\lambda &\left[ 
{\tilde f}(\lambda E_{\Delta,k}^+
- \alpha\delta\mu) \frac{\lambda E_{\Delta,k}^+ 
+ E_k^+}{2 E_{\Delta,k}^+} \frac{-p_0 + \lambda E_{\Delta,k}^+ 
+  E_{k'}^+}{(-p_0+\lambda E_{\Delta,k}^+)^2 
- (E^+_{\Delta,k'})^2} \right.\nonumber \\
&+& {\tilde f}(\lambda E_{\Delta,k'}^+
- \alpha\delta\mu) \frac{\lambda E_{\Delta,k'}^+ 
+ E_{k'}^+}{2 E_{\Delta,k'}^+} \frac{p_0 + \lambda E_{\Delta,k'}^+ 
+ E_{k}^+}{(p_0+\lambda E_{\Delta,k'}^+)^2 
- (E^+_{\Delta,k})^2} \nonumber \\
& + & {\tilde f}(\lambda E_{\Delta,k}^-
+ \alpha\delta\mu) \frac{\lambda E_{\Delta,k}^- 
+ E_k^-}{2 E_{\Delta,k}^-} \frac{-p_0 +\lambda E_{\Delta,k}^- 
+ E_{k'}^-}{(-p_0+\lambda E_{\Delta,k}^-)^2 
- (E^-_{\Delta,k'})^2} \nonumber \\
& + & \left.
{\tilde f}(\lambda E_{\Delta,k'}^-
+ \alpha\delta\mu) \frac{\lambda E_{\Delta,k'}^- +  E_{k'}^-}{2 E_{\Delta,k'}^-} 
\frac{p_0 + \lambda E_{\Delta,k'}^- +   E_{k}^-}{(p_0+ \lambda E_{\Delta,k'}^-)^2 -
(E^-_{\Delta,k})^2}\right], 
\end{eqnarray}

\begin{eqnarray}
C_{+-}^{11,22}=\sum_{\lambda=\pm}&\lambda &\left[ {\tilde f}(\lambda E_{\Delta,k}^+
+ \alpha\delta\mu) \frac{\lambda E_{\Delta,k}^+ -  E_k^+}{2 E_{\Delta,k}^+} 
\frac{-p_0 + \lambda E_{\Delta,k}^+ +   E_{k'}^-}{(-p_0+\lambda E_{\Delta,k}^+)^2 -
(E^-_{\Delta,k'})^2} \right. \nonumber \\
&+& {\tilde f}(\lambda E_{\Delta,k'}^+
-\alpha\delta\mu) \frac{\lambda E_{\Delta,k'}^+ +  E_{k'}^+}{2 E_{\Delta,k'}^+} 
\frac{p_0 + \lambda E_{\Delta,k'}^+ -   E_{k}^-}{(p_0+\lambda E_{\Delta,k'}^+)^2 -
(E^-_{\Delta,k})^2} \nonumber \\
& + & {\tilde f}(\lambda E_{\Delta,k}^-
- \alpha\delta\mu) \frac{\lambda E_{\Delta,k}^- -  E_k^-}{2 E_{\Delta,k}^-} 
\frac{-p_0 +\lambda E_{\Delta,k}^- +  E_{k'}^+}{(-p_0+\lambda E_{\Delta,k}^-)^2 -
(E^+_{\Delta,k'})^2} \nonumber \\
& + & \left. {\tilde f}(\lambda E_{\Delta,k'}^-
+ \alpha\delta\mu) \frac{\lambda E_{\Delta,k'}^- +  E_{k'}^-}{2 E_{\Delta,k'}^-} 
\frac{p_0 + \lambda E_{\Delta,k'}^- -   E_{k}^+}{(p_0+ \lambda E_{\Delta,k'}^-)^2 -
(E^+_{\Delta,k})^2}\right], 
\end{eqnarray}

\begin{eqnarray}
C_{-+}^{11,22}=\sum_{\lambda=\pm}&\lambda &\left[ {\tilde f}(\lambda E_{\Delta,k}^+
- \alpha\delta\mu) \frac{\lambda E_{\Delta,k}^+ +  E_k^+}{2 E_{\Delta,k}^+} 
\frac{-p_0 + \lambda E_{\Delta,k}^+ -  E_{k'}^-}{(-p_0+\lambda E_{\Delta,k}^+)^2 -
(E^-_{\Delta,k'})^2} \right. \nonumber \\
&+& {\tilde f}(\lambda E_{\Delta,k'}^+
+\alpha\delta\mu) \frac{\lambda E_{\Delta,k'}^+ -  E_{k'}^+}{2 E_{\Delta,k'}^+} 
\frac{p_0 + \lambda E_{\Delta,k'}^+ +  E_{k}^-}{(p_0+\lambda E_{\Delta,k'}^+)^2 -
(E^-_{\Delta,k})^2} \nonumber \\
& + & {\tilde f}(\lambda E_{\Delta,k}^-
+ \alpha\delta\mu) \frac{\lambda E_{\Delta,k}^- +  E_k^-}{2 E_{\Delta,k}^-} 
\frac{-p_0 +\lambda E_{\Delta,k}^- -  E_{k'}^+}{(-p_0+\lambda E_{\Delta,k}^-)^2 -
(E^+_{\Delta,k'})^2} \nonumber \\
& + & \left.
{\tilde f}(\lambda E_{\Delta,k'}^-
- \alpha\delta\mu) \frac{\lambda E_{\Delta,k'}^- -  E_{k'}^-}{2 E_{\Delta,k'}^-} 
\frac{p_0 + \lambda E_{\Delta,k'}^- +  E_{k}^+}{(p_0+ \lambda E_{\Delta,k'}^-)^2 -
(E^+_{\Delta,k})^2}\right], 
\end{eqnarray}

\begin{eqnarray}
C_{++}^{12,21}= - \Delta^2 \sum_{\lambda=\pm}&\lambda &
\left[ \frac{{\tilde f}(\lambda E_{\Delta,k}^+ - \alpha\delta\mu)}{2 E_{\Delta,k}^+}
\frac{1}{(-p_0+\lambda E_{\Delta,k}^+)^2 - (E^-_{\Delta,k'})^2} \right.\nonumber \\
&+& \frac{{\tilde f}(\lambda E_{\Delta,k'}^+ + \alpha\delta\mu)}{2 E_{\Delta,k'}^+} 
\frac{1}{(p_0+\lambda E_{\Delta,k'}^+)^2 - (E^-_{\Delta,k})^2} \nonumber \\
& + & \frac{{\tilde f}(\lambda E_{\Delta,k}^- + \alpha\delta\mu)}{2 E_{\Delta,k}^-} 
\frac{1}{(-p_0+\lambda E_{\Delta,k}^-)^2 - (E^+_{\Delta,k'})^2} \nonumber \\
& + & \left. \frac{{\tilde f}(\lambda E_{\Delta,k'}^- - \alpha\delta\mu)}{2 E_{\Delta,k'}^-} 
\frac{1}{(p_0+ \lambda E_{\Delta,k'}^-)^2 - (E^+_{\Delta,k})^2}\right], 
\end{eqnarray}

\begin{eqnarray}
C_{+-}^{12,21}= - \Delta^2 \sum_{\lambda=\pm}&\lambda &
\left[ \frac{{\tilde f}(\lambda E_{\Delta,k}^+ - \alpha\delta\mu)}{2 E_{\Delta,k}^+}
\frac{1}{(-p_0+\lambda E_{\Delta,k}^+)^2 - (E^+_{\Delta,k'})^2} \right.\nonumber \\
&+& \frac{{\tilde f}(\lambda E_{\Delta,k'}^+ - \alpha\delta\mu)}{2 E_{\Delta,k'}^+} 
\frac{1}{(p_0+\lambda E_{\Delta,k'}^+)^2 - (E^+_{\Delta,k})^2} \nonumber \\
& + & \frac{{\tilde f}(\lambda E_{\Delta,k}^- + \alpha\delta\mu)}{2 E_{\Delta,k}^-} 
\frac{1}{(-p_0+\lambda E_{\Delta,k}^-)^2 - (E^-_{\Delta,k'})^2} \nonumber \\
& + & \left. \frac{{\tilde f}(\lambda E_{\Delta,k'}^- + \alpha\delta\mu)}{2 E_{\Delta,k'}^-} 
\frac{1}{(p_0+ \lambda E_{\Delta,k'}^-)^2 - (E^-_{\Delta,k})^2}\right],
\end{eqnarray}
where ${\tilde f}(E)\equiv \left[1+\exp\left(E/T\right)\right]^{-1}$ is 
the Fermi distribution function. Note that
\begin{eqnarray}
& & C_{--}^{12}= C_{++}^{21}, \  \  
 C_{--}^{21}= C_{++}^{12}, \nonumber \\
& & C_{-+}^{12}= C_{+-}^{21}, \  \  
 C_{-+}^{21}= C_{+-}^{12}.
\end{eqnarray}

At zero temperature, these functions take the following form (here and 
below, the upper sign corresponds to the first case, e.g., $C_{++}^{11}$,
and the lower sign corresponds to the second case, e.g., $C_{++}^{22}$):
\begin{eqnarray}
C_{++}^{11,22} & = & 
 - \frac{(E_{\Delta,k}^- +  E_k^-)(E_{\Delta,k}^- +  E_{k'}^- +p_0)}
{2 E_{\Delta,k}^-  \left[(p_0 + E_{\Delta,k}^-)^2 - (E_{\Delta,k'}^-)^2\right]} 
 - \frac{(E_{\Delta,k'}^- +  E_{k'}^-)(E_{\Delta,k'}^- +  E_{k}^- - p_0)}
{2 E_{\Delta,k'}^-  \left[(p_0 - E_{\Delta,k'}^-)^2 - (E_{\Delta,k}^-)^2\right]} \nonumber \\
 &-& \frac{(E_{\Delta,k}^+ +  E_k^+)(E_{\Delta,k}^+ +  E_{k'}^+ +p_0)}
{2 E_{\Delta,k}^+  \left[(p_0 + E_{\Delta,k}^+)^2 - (E_{\Delta,k'}^+)^2\right]} 
 - \frac{(E_{\Delta,k'}^+ +  E_{k'}^+)(E_{\Delta,k'}^+ +  E_{k}^+ - p_0)}
{2 E_{\Delta,k'}^+  \left[(p_0 - E_{\Delta,k'}^+)^2 - (E_{\Delta,k}^+)^2\right]} \nonumber \\
& + & \theta(-E_{\Delta,k}^-+\delta\mu) \frac{(E_{\Delta,k}^- \mp  E_k^-)
(E_{\Delta,k}^- \mp  E_{k'}^- \mp p_0)}
{2 E_{\Delta,k}^- \left[(p_0 \mp E_{\Delta,k}^-)^2 - (E_{\Delta,k'}^-)^2\right]} \nonumber \\
& + & \theta(-E_{\Delta,k'}^-+\delta\mu) \frac{(E_{\Delta,k'}^- \mp  E_{k'}^-)
(E_{\Delta,k'}^- \mp  E_{k}^- \pm p_0)}
{2 E_{\Delta,k'}^- \left[(p_0 \pm E_{\Delta,k'}^-)^2 - (E_{\Delta,k}^-)^2\right]},
\end{eqnarray}

\begin{eqnarray}
C_{--}^{11,22} & = & 
 - \frac{(E_{\Delta,k}^- -  E_k^-)(E_{\Delta,k}^- -  E_{k'}^- + p_0)}
{2 E_{\Delta,k}^-  \left[(p_0 + E_{\Delta,k}^-)^2 - (E_{\Delta,k'}^-)^2\right]} 
 - \frac{(E_{\Delta,k'}^- -  E_{k'}^-)(E_{\Delta,k'}^- -  E_{k}^- - p_0)}
{2 E_{\Delta,k'}^-  \left[(p_0 - E_{\Delta,k'}^-)^2 - (E_{\Delta,k}^-)^2\right]} \nonumber \\
 &-& \frac{(E_{\Delta,k}^+ -  E_k^+)(E_{\Delta,k}^+ -  E_{k'}^+ +p_0)}
{2 E_{\Delta,k}^+  \left[(p_0 + E_{\Delta,k}^+)^2 - (E_{\Delta,k'}^+)^2\right]} 
 - \frac{(E_{\Delta,k'}^+ -  E_{k'}^+)(E_{\Delta,k'}^+ -  E_{k}^+ - p_0)}
{2 E_{\Delta,k'}^+  \left[(p_0 - E_{\Delta,k'}^+)^2 - (E_{\Delta,k}^+)^2\right]} \nonumber \\
& + & \theta(-E_{\Delta,k}^-+\delta\mu) \frac{(E_{\Delta,k}^- \mp  E_k^-)
(E_{\Delta,k}^- \mp  E_{k'}^- \pm p_0)}
{2 E_{\Delta,k}^- \left[(p_0 \pm E_{\Delta,k}^-)^2 - (E_{\Delta,k'}^-)^2\right]} \nonumber \\
& + & \theta(-E_{\Delta,k'}^-+\delta\mu) \frac{(E_{\Delta,k'}^- \mp  E_{k'}^-)
(E_{\Delta,k'}^- \mp  E_{k}^- \mp p_0)}
{2 E_{\Delta,k'}^- \left[(p_0 \mp E_{\Delta,k'}^-)^2 - (E_{\Delta,k}^-)^2\right]}, 
\end{eqnarray}

\begin{eqnarray}
C_{+-}^{11,22} & = & 
 - \frac{(E_{\Delta,k}^- +  E_k^-)(E_{\Delta,k}^- -  E_{k'}^+ +p_0)}
{2 E_{\Delta,k}^-  \left[(p_0 + E_{\Delta,k}^-)^2 - (E_{\Delta,k'}^+)^2\right]} 
 - \frac{(E_{\Delta,k'}^- -  E_{k'}^-)(E_{\Delta,k'}^- +  E_{k}^+ - p_0)}
{2 E_{\Delta,k'}^-  \left[(p_0 - E_{\Delta,k'}^-)^2 - (E_{\Delta,k}^+)^2\right]} \nonumber \\
 &-& \frac{(E_{\Delta,k}^+ +  E_k^+)(E_{\Delta,k}^+ -  E_{k'}^- +p_0)}
{2 E_{\Delta,k}^+  \left[(p_0 + E_{\Delta,k}^+)^2 - (E_{\Delta,k'}^-)^2\right]} 
 - \frac{(E_{\Delta,k'}^+ -  E_{k'}^+)(E_{\Delta,k'}^+ +  E_{k}^- - p_0)}
{2 E_{\Delta,k'}^+  \left[(p_0 - E_{\Delta,k'}^+)^2 - (E_{\Delta,k}^-)^2\right]} \nonumber \\
& + & \theta(-E_{\Delta,k}^-+\delta\mu) \frac{(E_{\Delta,k}^- \mp  E_k^-)
(E_{\Delta,k}^- \pm  E_{k'}^+ \mp p_0)}
{2 E_{\Delta,k}^- \left[(p_0 \mp E_{\Delta,k}^-)^2 - (E_{\Delta,k'}^+)^2\right]} \nonumber \\
& + & \theta(-E_{\Delta,k'}^-+\delta\mu) \frac{(E_{\Delta,k'}^- \mp  E_{k'}^-)
(E_{\Delta,k'}^- \pm  E_{k}^+ \mp p_0)}
{2 E_{\Delta,k'}^- \left[(p_0 \mp E_{\Delta,k'}^-)^2 - (E_{\Delta,k}^+)^2\right]}, 
\end{eqnarray}

\begin{eqnarray}
C_{-+}^{11,22} & = & 
 - \frac{(E_{\Delta,k}^- -  E_k^-)(E_{\Delta,k}^- +  E_{k'}^+ +p_0)}
{2 E_{\Delta,k}^-  \left[(p_0 + E_{\Delta,k}^-)^2 - (E_{\Delta,k'}^+)^2\right]} 
 - \frac{(E_{\Delta,k'}^- +  E_{k'}^-)(E_{\Delta,k'}^- -  E_{k}^+ - p_0)}
{2 E_{\Delta,k'}^-  \left[(p_0 - E_{\Delta,k'}^-)^2 - (E_{\Delta,k}^+)^2\right]} \nonumber \\
 &-& \frac{(E_{\Delta,k}^+ -  E_k^+)(E_{\Delta,k}^+ +  E_{k'}^- +p_0)}
{2 E_{\Delta,k}^+  \left[(p_0 + E_{\Delta,k}^+)^2 - (E_{\Delta,k'}^-)^2\right]} 
 - \frac{(E_{\Delta,k'}^+ +  E_{k'}^+)(E_{\Delta,k'}^+ -  E_{k}^- - p_0)}
{2 E_{\Delta,k'}^+  \left[(p_0 - E_{\Delta,k'}^+)^2 - (E_{\Delta,k}^-)^2\right]} \nonumber \\
& + & \theta(-E_{\Delta,k}^-+\delta\mu) \frac{(E_{\Delta,k}^- \mp  E_k^-)
(E_{\Delta,k}^- \pm  E_{k'}^+ \pm p_0)}
{2 E_{\Delta,k}^- \left[(p_0 \pm E_{\Delta,k}^-)^2 - (E_{\Delta,k'}^+)^2\right]} \nonumber \\
& + & \theta(-E_{\Delta,k'}^-+\delta\mu) \frac{(E_{\Delta,k'}^- \mp  E_{k'}^-)
(E_{\Delta,k'}^- \pm  E_{k}^+ \pm p_0)}
{2 E_{\Delta,k'}^- \left[(p_0 \pm E_{\Delta,k'}^-)^2 - (E_{\Delta,k}^+)^2\right]}, 
\end{eqnarray}

\begin{eqnarray}
C_{++}^{12,21} & = & C_{--}^{12,21}=
 - \Delta^2 \Bigg[
 - \frac{1}{2E_{\Delta,k}^-\left[(p_0+E_{\Delta,k}^-)^2 - (E_{\Delta,k'}^+)^2\right]} - 
\frac{1}{2E_{\Delta,k'}^-\left[(p_0-E_{\Delta,k'}^-)^2 - (E_{\Delta,k}^+)^2\right]} \nonumber \\
&-& \frac{1}{2E_{\Delta,k}^+\left[(p_0+E_{\Delta,k}^+)^2 - (E_{\Delta,k'}^-)^2\right]} - 
\frac{1}{2E_{\Delta,k'}^+\left[(p_0-E_{\Delta,k'}^+)^2 - (E_{\Delta,k}^-)^2\right]} \nonumber \\
& + & \frac{\theta(-E_{\Delta,k}^-+\delta\mu)}{2E_{\Delta,k}^-}
\frac{1}{(p_0 \pm E_{\Delta,k}^-)^2-(E_{\Delta,k'}^+)^2} 
+ \frac{\theta(-E_{\Delta,k'}^-+\delta\mu)}{2E_{\Delta,k'}^-}
\frac{1}{(p_0 \pm E_{\Delta,k'}^-)^2-(E_{\Delta,k}^+)^2}\Bigg],
\end{eqnarray}

\begin{eqnarray}
C_{+-}^{12,21} & = & C_{-+}^{12,21}=
 - \Delta^2\Bigg[
 - \frac{1}{2E_{\Delta,k}^-\left[(p_0+E_{\Delta,k}^-)^2 - (E_{\Delta,k'}^-)^2\right]} - 
\frac{1}{2E_{\Delta,k'}^-\left[(p_0-E_{\Delta,k'}^-)^2 - (E_{\Delta,k}^-)^2\right]} \nonumber \\
&-& \frac{1}{2E_{\Delta,k}^+\left[(p_0+E_{\Delta,k}^+)^2 - (E_{\Delta,k'}^+)^2\right]} - 
\frac{1}{2E_{\Delta,k'}^+\left[(p_0-E_{\Delta,k'}^+)^2 - (E_{\Delta,k}^+)^2\right]} \nonumber \\
& + & \frac{\theta(-E_{\Delta,k}^-+\delta\mu)}{2E_{\Delta,k}^-}
\frac{1}{(p_0 \pm E_{\Delta,k}^-)^2-(E_{\Delta,k'}^-)^2} 
+ \frac{\theta(-E_{\Delta,k'}^-+\delta\mu)}{2E_{\Delta,k'}^-}
\frac{1}{(p_0 \pm E_{\Delta,k'}^-)^2-(E_{\Delta,k}^-)^2}\Bigg].
\end{eqnarray}

\subsection{The coefficient functions for $\Pi^{AB}_{\mu\nu}$ 
with $A,B=4,5,6,7$}
\label{Cof-TU44}

The components of the polarization tensor $\Pi^{AB}_{\mu\nu}$ with 
$A,B=4,5,6,7$ are given in terms of the following coefficient 
functions:
\begin{eqnarray}
C_{++}^{44} & = & \sum_{\lambda,\alpha=\pm} \lambda  
\left[\frac{{\tilde f}(\lambda E_{\Delta,k'}^+ + \alpha \delta\mu)}{2 E_{\Delta,k'}^+ }
\frac{\lambda E_{\Delta,k'}^+ -  E_{k'}^+}{p_0+ \lambda E_{\Delta,k'}^+ 
+  E_{b,k}^+} \right. \nonumber \\
& + &\left. 
\frac{{\tilde f}(\lambda E_{\Delta,k}^- + \alpha \delta\mu)}{2 E_{\Delta,k}^- }
\frac{\lambda E_{\Delta,k}^- -  E_{k}^-}{-p_0+ \lambda E_{\Delta,k}^- 
+  E_{b,k'}^-}\right]\nonumber \\
& + &\left[{\tilde f}(- E_{bu,k}^+) + {\tilde f}(- E_{bd,k}^+)\right]\frac{-p_0- E_{b,k}^+ -  E_{k'}^+}
{(p_0+ E_{b,k}^+)^2-({E_{\Delta,k'}^+})^2} \nonumber \\
& + &\left[{\tilde f}(- E_{bu,k'}^-) + {\tilde f}(- E_{bd,k'}^-)\right]\frac{p_0- E_{b,k'}^- -  E_{k}^-}
{(p_0- E_{b,k'}^-)^2-(E^-_{\Delta,k})^2},
\end{eqnarray}

\begin{eqnarray}
C_{--}^{44} & = & \sum_{\lambda,\alpha=\pm} \lambda 
\left[\frac{{\tilde f}(\lambda E_{\Delta,k}^+ + \alpha \delta\mu)}{2 E_{\Delta,k}^+ }
\frac{\lambda E_{\Delta,k}^+ +  E_{k}^+}{-p_0+ \lambda E_{\Delta,k}^+ 
-  E_{b,k'}^+} \right.\nonumber \\
& + &\left.
\frac{{\tilde f}(\lambda E_{\Delta,k'}^- + \alpha \delta\mu)}{2 E_{\Delta,k'}^- }
\frac{\lambda E_{\Delta,k'}^- +  E_{k'}^-}{p_0+ \lambda E_{\Delta,k'}^- 
-  E_{b,k}^-}\right]\nonumber \\
& + &\left[{\tilde f}(E_{bu,k}^-) + {\tilde f}(E_{bd,k}^-)\right]\frac{-p_0+ E_{b,k}^- +  E_{k'}^-}
{(p_0- E_{b,k}^-)^2-(E_{\Delta,k'}^-)^2} \nonumber \\
& + &\left[{\tilde f}(E_{bu,k'}^+) + {\tilde f}(- E_{bd,k'}^+)\right]\frac{p_0+ E_{b,k'}^+ +  E_{k}^+}
{(p_0+ E_{b,k'}^+)^2-(E^+_{\Delta,k})^2},
\end{eqnarray}

\begin{eqnarray}
C_{+-}^{44} & = & \sum_{\lambda,\alpha=\pm} \lambda 
\left[\frac{{\tilde f}(\lambda E_{\Delta,k}^- + \alpha \delta\mu)}{2 E_{\Delta,k}^-}
\frac{\lambda E_{\Delta,k}^- -  E_{k}^-}{-p_0+ \lambda E_{\Delta,k}^- 
-  E_{b,k'}^+} \right.\nonumber \\
& + &\left.
\frac{{\tilde f}(\lambda E_{\Delta,k'}^- + \alpha \delta\mu)}{2 E_{\Delta,k'}^- }
\frac{\lambda E_{\Delta,k'}^- +  E_{k'}^-}{p_0+ \lambda E_{\Delta,k'}^- 
+  E_{b,k}^+}\right]\nonumber \\
& + &\left[{\tilde f}(-E_{bu,k}^+) + {\tilde f}(-E_{bd,k}^+)\right]\frac{-p_0- E_{b,k}^+ +  E_{k'}^-}
{(p_0+ E_{b,k}^+)^2-(E_{\Delta,k'}^-)^2} \nonumber \\
& + &\left[{\tilde f}(E_{bu,k'}^+) + {\tilde f}( E_{bd,k'}^+)\right]\frac{p_0+ E_{b,k'}^+ -  E_{k}^-}
{(p_0+ E_{b,k'}^+)^2-(E^-_{\Delta,k})^2}, 
\end{eqnarray}

\begin{eqnarray}
C_{-+}^{44} & = & \sum_{\lambda,\alpha=\pm} \lambda 
\left[\frac{{\tilde f}(\lambda E_{\Delta,k}^+ + \alpha \delta\mu)}{2 E_{\Delta,k}^+}
\frac{\lambda E_{\Delta,k}^+ +  E_{k}^+}{-p_0+ \lambda E_{\Delta,k}^+ 
+  E_{b,k'}^-} \right.\nonumber \\
& + &\left.
\frac{{\tilde f}(\lambda E_{\Delta,k'}^+ + \alpha \delta\mu)}{2 E_{\Delta,k'}^+ }
\frac{\lambda E_{\Delta,k'}^+ -  E_{k'}^+}{p_0+ \lambda E_{\Delta,k'}^+ 
-  E_{b,k}^-}\right]\nonumber \\
& + &\left[{\tilde f}(E_{bu,k}^-) + {\tilde f}(E_{bd,k}^-)\right]\frac{-p_0+ E_{b,k}^- -  E_{k'}^+}
{(p_0- E_{b,k}^-)^2-(E_{\Delta,k'}^+)^2} \nonumber \\
& + &\left[{\tilde f}(-E_{bu,k'}^-) + {\tilde f}(- E_{bd,k'}^-)\right]\frac{p_0- E_{b,k'}^- +  E_{k}^+}
{(p_0- E_{b,k'}^-)^2-(E^+_{\Delta,k})^2},
\end{eqnarray}

\begin{eqnarray}
C_{++}^{55} & = & \sum_{\lambda,\alpha=\pm} \lambda 
\left[\frac{{\tilde f}(\lambda E_{\Delta,k}^+ + \alpha \delta\mu)}{2 E_{\Delta,k}^+ }
\frac{\lambda E_{\Delta,k}^+ -  E_{k}^+}{-p_0+ \lambda E_{\Delta,k}^+ 
+  E_{b,k'}^+} \right.\nonumber \\
& + &\left.
\frac{{\tilde f}(\lambda E_{\Delta,k'}^- + \alpha \delta\mu)}{2 E_{\Delta,k'}^- }
\frac{\lambda E_{\Delta,k'}^- -  E_{k'}^-}{p_0+ \lambda E_{\Delta,k'}^- 
+  E_{b,k}^-}\right]\nonumber \\
& + &\left[{\tilde f}(- E_{bu,k'}^+) + {\tilde f}(- E_{bd,k'}^+)\right]
\frac{p_0- E_{b,k'}^+ -  E_{k}^+}
{(p_0- E_{b,k'}^+)^2-({E_{\Delta,k}^+})^2} \nonumber \\
& + &\left[{\tilde f}(- E_{bu,k}^-) + {\tilde f}(- E_{bd,k}^-)\right]
\frac{-p_0- E_{b,k}^- -  E_{k'}^-}
{(p_0+ E_{b,k}^-)^2-(E^-_{\Delta,k'})^2},
\end{eqnarray}

\begin{eqnarray}
C_{--}^{55} & = & \sum_{\lambda,\alpha=\pm} \lambda 
\left[\frac{{\tilde f}(\lambda E_{\Delta,k'}^+ + \alpha \delta\mu)}{2 E_{\Delta,k'}^+ }
\frac{\lambda E_{\Delta,k'}^+ +  E_{k'}^+}{p_0+ \lambda E_{\Delta,k'}^+ 
-  E_{b,k}^+} \right.\nonumber \\
& + &\left.
\frac{{\tilde f}(\lambda E_{\Delta,k}^- + \alpha \delta\mu)}{2 E_{\Delta,k}^- }
\frac{\lambda E_{\Delta,k}^- +  E_{k}^-}{-p_0+ \lambda E_{\Delta,k}^- 
-  E_{b,k'}^-}\right]\nonumber \\
& + &\left[{\tilde f}(E_{bu,k'}^-) + {\tilde f}(E_{bd,k'}^-)\right]\frac{p_0+ E_{b,k'}^- +  E_{k}^-}
{(p_0+ E_{b,k'}^-)^2-(E_{\Delta,k}^-)^2} \nonumber \\
& + &\left[{\tilde f}(E_{bu,k}^+) + {\tilde f}(- E_{bd,k}^+)\right]\frac{-p_0+ E_{b,k}^+ +  E_{k'}^+}
{(p_0- E_{b,k}^+)^2-(E^+_{\Delta,k'})^2}, 
\end{eqnarray}

\begin{eqnarray}
C_{+-}^{55} & = & \sum_{\lambda,\alpha=\pm} \lambda 
\left[\frac{{\tilde f}(\lambda E_{\Delta,k}^+ + \alpha \delta\mu)}{2 E_{\Delta,k}^+}
\frac{\lambda E_{\Delta,k}^+ -  E_{k}^+}{-p_0+ \lambda E_{\Delta,k}^+ 
-  E_{b,k'}^-} \right.\nonumber \\
& + &\left.
\frac{{\tilde f}(\lambda E_{\Delta,k'}^+ + \alpha \delta\mu)}{2 E_{\Delta,k'}^+ }
\frac{\lambda E_{\Delta,k'}^+ +  E_{k'}^+}{p_0+ \lambda E_{\Delta,k'}^+ 
+  E_{b,k}^-}\right]\nonumber \\
& + &\left[{\tilde f}(-E_{bu,k}^-) + {\tilde f}(-E_{bd,k}^-)\right]\frac{-p_0- E_{b,k}^- +  E_{k'}^+}
{(p_0+ E_{b,k}^-)^2-(E_{\Delta,k'}^+)^2} \nonumber \\
& + &\left[{\tilde f}(E_{bu,k'}^-) + {\tilde f}(E_{bd,k'}^-)\right]\frac{p_0+ E_{b,k'}^- - E_{k}^+}
{(p_0+ E_{b,k'}^-)^2-(E^+_{\Delta,k})^2}, 
\end{eqnarray}

\begin{eqnarray}
C_{-+}^{55} & = & \sum_{\lambda,\alpha=\pm} \lambda 
\left[\frac{{\tilde f}(\lambda E_{\Delta,k}^- + \alpha \delta\mu)}{2 E_{\Delta,k}^-}
\frac{\lambda E_{\Delta,k}^- +  E_{k}^-}{-p_0+ \lambda E_{\Delta,k}^- 
+  E_{b,k'}^+} \right.\nonumber \\
& + &\left.
\frac{{\tilde f}(\lambda E_{\Delta,k'}^- + \alpha \delta\mu)}{2 E_{\Delta,k'}^- }
\frac{\lambda E_{\Delta,k'}^- -  E_{k'}^-}{p_0+ \lambda E_{\Delta,k'}^- 
-  E_{b,k}^+}\right]\nonumber \\
& + &\left[{\tilde f}(E_{bu,k}^+) + {\tilde f}(E_{bd,k}^+)\right]\frac{-p_0+ E_{b,k}^+ -  E_{k'}^-}
{(p_0- E_{b,k}^+)^2-(E_{\Delta,k'}^-)^2} \nonumber \\
& + &\left[{\tilde f}(-E_{bu,k'}^+) + {\tilde f}(-E_{bd,k'}^+)\right]\frac{p_0- E_{b,k'}^+ +  E_{k}^-}
{(p_0- E_{b,k'}^+)^2-(E^-_{\Delta,k})^2}. 
\end{eqnarray}


At zero temperature, the expressions for these coefficients become 
\begin{eqnarray}
C_{++}^{44} &=&- \frac{1}{E_{\Delta,k'}^+}
\frac{E_{\Delta,k'}^+- E_{k'}^+}{E_{\Delta,k'}^++  E_{b,k}^+ +p_0}
-\frac{1}{E_{\Delta,k}^-}
\frac{E_{\Delta,k}^-- E_{k}^-}{E_{\Delta,k}^- +  E_{b,k'}^- -p_0} \nonumber \\
&+& \frac{\theta(-E_{\Delta,k}^-+\delta\mu)}{E_{\Delta,k}^-}
\frac{(E_{\Delta,k}^-)^2 + E_{k}^-( E_{b,k'}^- -p_0)}
{(E_{\Delta,k}^-)^2- ( E_{b,k'}^- -p_0)^2}
 \nonumber \\
&-&\left[ \theta(-E_{bu,k'}^-)+\theta(-E_{bd,k'}^-)\right] 
\frac{ E_{b,k'}^- + E_{k}^- -p_0}{(E_{\Delta,k}^-)^2- ( E_{b,k'}^- -p_0)^2},
\end{eqnarray}

\begin{eqnarray}
C_{--}^{44} &=& -\frac{1}{E_{\Delta,k}^+}
\frac{E_{\Delta,k}^+- E_{k}^+}{E_{\Delta,k}^+ +  E_{b,k'}^+ + p_0}
-\frac{1}{E_{\Delta,k'}^-}
\frac{E_{\Delta,k'}^- - E_{k'}^-}{E_{\Delta,k'}^- +  E_{b,k}^- - p_0} \nonumber \\
&+& \frac{\theta(-E_{\Delta,k'}^-+\delta\mu)}{E_{\Delta,k'}^-}
\frac{(E_{\Delta,k'}^-)^2 + E_{k'}^-( E_{b,k}^- -p_0)}
{(E_{\Delta,k'}^-)^2- ( E_{b,k}^- -p_0)^2} \nonumber \\
&-&\left[ \theta(-E_{bu,k}^-)+\theta(-E_{bd,k}^-)\right] 
\frac{ E_{b,k}^- + E_{k'}^- -p_0}{(E_{\Delta,k'}^-)^2- ( E_{b,k'}^- -p_0)^2},
\end{eqnarray}

\begin{eqnarray}
C_{+-}^{44} &=&-\frac{1}{E_{\Delta,k}^-}
\frac{E_{\Delta,k}^- + E_{k}^-}{E_{\Delta,k}^-+  E_{b,k'}^+ +p_0}
-\frac{1}{E_{\Delta,k'}^-}
\frac{E_{\Delta,k'}^-+ E_{k'}^-}{E_{\Delta,k'}^- +  E_{b,k}^+ +p_0} 
\nonumber \\
&+& \frac{\theta(-E_{\Delta,k'}^-+\delta\mu)}{E_{\Delta,k'}^-}
\frac{(E_{\Delta,k'}^-)^2-  E_{k'}^- ( E_{b,k}^+ + p_0)}
{(E_{\Delta,k'}^-)^2-( E_{b,k}^+ + p_0)^2} \nonumber \\
& + & \frac{\theta(-E_{\Delta,k}^-+\delta\mu)}{E_{\Delta,k}^-}
\frac{(E_{\Delta,k}^-)^2-  E_{k}^- ( E_{b,k'}^+ + p_0)}
{(E_{\Delta,k}^-)^2-( E_{b,k'}^+ + p_0)^2},
\end{eqnarray}

\begin{eqnarray}
C_{-+}^{44} &=&-\frac{1}{E_{\Delta,k}^+}
\frac{E_{\Delta,k}^+ + E_{k}^+}{E_{\Delta,k}^++  E_{b,k'}^- -p_0}
-\frac{1}{E_{\Delta,k'}^+}
\frac{E_{\Delta,k'}^++ E_{k'}^+}{E_{\Delta,k'}^+ +  E_{b,k}^- -p_0} \nonumber \\
&-&\left[ \theta(-E_{bu,k}^-) + \theta(-E_{bd,k}^-)\right] 
\frac{ E_{b,k}^- - E_{k'}^+ -p_0}{(E_{\Delta,k'}^+)^2-( E_{b,k}^- -p_0)^2} \nonumber \\
&-&\left[ \theta(-E_{bu,k'}^-) + \theta(-E_{bd,k'}^-)\right] 
\frac{ E_{b,k'}^- - E_{k}^+ -p_0}{(E^+_{\Delta,k})^2-( E_{b,k'}^- -p_0)^2},
\end{eqnarray}

\begin{eqnarray}
C_{++}^{55} &=&- \frac{1}{E_{\Delta,k}^+}
\frac{E_{\Delta,k}^+- E_{k}^+}{E_{\Delta,k}^++  E_{b,k'}^+ -p_0}
-\frac{1}{E_{\Delta,k'}^-}
\frac{E_{\Delta,k'}^-- E_{k'}^-}{E_{\Delta,k'}^- +  E_{b,k}^- +p_0} \nonumber \\
&+& \frac{\theta(-E_{\Delta,k'}^-+\delta\mu)}{E_{\Delta,k'}^-}
\frac{(E_{\Delta,k'}^-)^2 + E_{k'}^-( E_{b,k}^- + p_0)}
{(E_{\Delta,k'}^-)^2- ( E_{b,k}^- +p_0)^2}
 \nonumber \\
&-&\left[ \theta(-E_{bu,k}^-)+\theta(-E_{bd,k}^-)\right] 
\frac{ E_{b,k}^- + E_{k'}^- +p_0}{(E_{\Delta,k'}^-)^2- ( E_{b,k}^- +p_0)^2},
\end{eqnarray}


\begin{eqnarray}
C_{--}^{55} &=& -\frac{1}{E_{\Delta,k'}^+}
\frac{E_{\Delta,k'}^+- E_{k'}^+}{E_{\Delta,k'}^++  E_{b,k}^+ -p_0}
-\frac{1}{E_{\Delta,k}^-}
\frac{E_{\Delta,k}^-- E_{k}^-}{E_{\Delta,k}^- +  E_{b,k'}^- +p_0} \nonumber \\
&+& \frac{\theta(-E_{\Delta,k}^-+\delta\mu)}{E_{\Delta,k}^-}
\frac{(E_{\Delta,k}^-)^2 + E_{k}^-( E_{b,k'}^- +p_0)}
{(E_{\Delta,k}^-)^2- ( E_{b,k'}^- +p_0)^2} \nonumber \\
&-&\left[ \theta(-E_{bu,k'}^-)+\theta(-E_{bd,k'}^-)\right] 
\frac{ E_{b,k'}^- + E_{k}^- +p_0}{(E_{\Delta,k}^-)^2- ( E_{b,k'}^- +p_0)^2},
\end{eqnarray}


\begin{eqnarray}
C_{+-}^{55} &=&-\frac{1}{E_{\Delta,k}^+}
\frac{E_{\Delta,k}^+ + E_{k}^+}{E_{\Delta,k}^++  E_{b,k'}^- +p_0}
-\frac{1}{E_{\Delta,k'}^+}
\frac{E_{\Delta,k'}^++ E_{k'}^+}{E_{\Delta,k'}^+ +  E_{b,k}^- +p_0} \nonumber \\
&-&\left[ \theta(-E_{bu,k}^-) + \theta(-E_{bd,k}^-)\right] 
\frac{ E_{b,k}^- - E_{k'}^+ +p_0}{(E_{\Delta,k'}^+)^2-( E_{b,k}^- +p_0)^2} \nonumber \\
&-&\left[ \theta(-E_{bu,k'}^-) + \theta(-E_{bd,k'}^-)\right] 
\frac{ E_{b,k'}^- - E_{k}^+ +p_0}{(E^+_{\Delta,k})^2-( E_{b,k'}^- +p_0)^2},
\end{eqnarray}


\begin{eqnarray}
C_{-+}^{55} &=&-\frac{1}{E_{\Delta,k}^-}
\frac{E_{\Delta,k}^- + E_{k}^-}{E_{\Delta,k}^-+  E_{b,k'}^+ -p_0}
-\frac{1}{E_{\Delta,k'}^-}
\frac{E_{\Delta,k'}^-+ E_{k'}^-}{E_{\Delta,k'}^- +  E_{b,k}^+ -p_0} \nonumber \\
&+& \frac{\theta(-E_{\Delta,k'}^-+\delta\mu)}{E_{\Delta,k'}^-}
\frac{(E_{\Delta,k'}^-)^2-  E_{k'}^- ( E_{b,k}^+ - p_0)}
{(E_{\Delta,k'}^-)^2-( E_{b,k}^+ - p_0)^2} \nonumber \\
& + & \frac{\theta(-E_{\Delta,k}^-+\delta\mu)}{E_{\Delta,k}^-}
\frac{(E_{\Delta,k}^-)^2-  E_{k}^- ( E_{b,k'}^+ - p_0)}
{(E_{\Delta,k}^-)^2-( E_{b,k'}^+ - p_0)^2}.
\end{eqnarray}


\section{Momentum integrals}

\subsection{Polarization tensor $\Pi^{AB}_{\mu\nu}$ 
with $A,B=1,2,3,8,9$}
\label{Int-11}

In momentum integrals that appear in the expressions for the 
polarization tensor $\Pi^{AB}_{\mu\nu}$ 
with $A,B=1,2,3,8,9$, one can use some useful approximations 
that simplify the calculations considerably. For example, in 
all integrals that come from particle-hole loops, the main 
contribution comes from a close vicinity of the average quark 
Fermi momentum $p\simeq\bar\mu$. In these integrals, therefore, 
it is justified to make the following replacement:
\begin{equation}
\int_0^{\infty} d k k^2 (\ldots)_{\mbox{\scriptsize p-h}} \simeq
\bar\mu^2 \int_0^{\infty} dk (\ldots)_{\mbox{\scriptsize p-h}}.
\end{equation}
The corrections to such momentum integrals are suppressed by inverse
powers of $\bar\mu$. To be consistent with this approximation, the 
antiparticle-antiparticle loops should be omitted, and the dependence 
of the particle-antiparticle loops on $\Delta$, $\delta\mu$ and $\mu_8$ 
may be neglected. As a result, the only type of the particle-antiparticle 
contribution, that appears in the calculation, has the form
\begin{equation}
\int_0^{\infty} d k k^2\left(-\frac{|k - \bar\mu | + k - \bar\mu }
{ |k - \bar\mu | (k + \bar\mu + |k - \bar\mu |)} +\frac{1}{k}\right)
= \int_0^{\mu} k d k =\frac{\bar\mu^2}{2}.
\end{equation}
The particle-hole loops give rise to the integrals of the following type:

\begin{eqnarray}
\int_0^{\infty} dk k^2  ~ \frac{\Delta^2}{(E_{\Delta,k}^-)^3}  & \simeq  & 
2 ~ {\bar \mu}^2,\\
\int_0^{\infty} dk k^2  ~ \frac{ E_k^-}{E_{\Delta,k}^-} 
\delta(-E_{\Delta, k}^- + \delta\mu)  & \simeq  & 0,
\label{simeq0}\\
\int_0^{\infty} d k k^2  ~ \frac{\Delta^2}{(E_{\Delta,k}^-)^3} 
\theta(-E_{\Delta, k}^- + \delta\mu) 
& \simeq & 2 ~ {\bar \mu}^2 \frac{\sqrt{(\delta\mu)^2-\Delta^2}}{\delta\mu} 
\theta(\delta\mu-\Delta),\\
\int_0^{\infty} dk k^2  ~ \frac{\Delta^2}{(E_{\Delta,k}^-)^2} 
\delta(-E_{\Delta, k}^- + \delta\mu) 
& = &  2 \bar\mu^2 
\frac{\Delta^2}{\delta\mu\sqrt{(\delta\mu)^2-\Delta^2}}\theta(\delta\mu-\Delta),\\
\int_0^{\infty} dk k^2  ~ \frac{(E_{\Delta,k}^-)^2+( E_k^-)^2}{(E_{\Delta,k}^-)^2} 
\delta(-E_{\Delta, k}^- + \delta\mu) & = & 2 \bar\mu^2
\Bigg[ \frac{\delta\mu}{\sqrt{(\delta\mu)^2-\Delta^2}} 
  + \frac{\sqrt{(\delta\mu)^2-\Delta^2}}{\delta\mu} \Bigg]\theta(\delta\mu-\Delta) . 
\end{eqnarray}

\subsection{Polarization tensor $\Pi^{AB}_{\mu\nu}$ 
with A,B=4,5,6,7}
\label{Int-44}

The calculation of the polarization tensor $\Pi^{AB}_{\mu\nu}$ 
with $A,B=4,5,6,7$ reduces to the calculation of the integral
in Eq.~(\ref{c4pp-integral}). This can be written as a sum of 
two expressions. One of them comes from integrating the first 
line in Eq.~(\ref{cccc}). The result is
\begin{eqnarray}
&&\int_{0}^{\mu_{ub}} k^2dk
\frac{E_{\Delta,k}^{-}+k-\bar\mu}
{E_{\Delta,k}^{-}(E_{\Delta,k}^{-}-k+\bar\mu_b)}
+ \int_{\mu_{ub}}^{\mu_{db}} k^2dk
\frac{(E_{\Delta,k}^{-})^2+(k-\bar\mu)(k-\bar\mu_b)}
{E_{\Delta,k}^{-}\left[(E_{\Delta,k}^{-})^2-(k-\bar\mu_b)^2\right]}
\nonumber\\
&& + \int_{\mu_{db}}^{\infty} k^2dk
\frac{E_{\Delta,k}^{-}-k+\bar\mu}
{E_{\Delta,k}^{-}(E_{\Delta,k}^{-}+k-\bar\mu_b)}
\simeq \bar\mu^2 \left(
\int_{-\bar\mu}^{-\delta\mu-\mu_8} 
\frac{dx(\sqrt{x^2+\Delta^2}+x)}{\sqrt{x^2+\Delta^2}
(\sqrt{x^2+\Delta^2}-x-\mu_8)}\right. \nonumber\\
&&\left. +\int_{-\delta\mu-\mu_8}^{\delta\mu-\mu_8} 
\frac{dx(2x^2+\Delta^2+x\mu_8)}{\sqrt{x^2+\Delta^2}
(\Delta^2 - 2 x \mu_8 - \mu_8^2)}
+\int_{\delta\mu-\mu_8}^{\infty} 
\frac{dx(\sqrt{x^2+\Delta^2}-x)}{\sqrt{x^2+\Delta^2}
(\sqrt{x^2+\Delta^2}+x+\mu_8)}\right)\nonumber\\
&\simeq & 2\bar\mu^2\left[
1-\frac{\Delta^2}{4\mu_8^2}\ln 
\frac{\left(\Delta^2+\mu_8^2\right)^2-\left(\mu_e\mu_8\right)^2}
{\Delta^4}\right],
\end{eqnarray}
where $\bar\mu_b=\bar\mu-\mu_8$ and $x=k-\bar\mu$. In this calculation, 
we used the 
following table integrals:
\begin{eqnarray}
\int \frac{dx(\sqrt{x^2+\Delta^2}+x)}{\sqrt{x^2+\Delta^2}
(\sqrt{x^2+\Delta^2}-x-\mu_8)} &=& 
-\frac{\sqrt{x^2+\Delta^2}+x}{\mu_8}
-\frac{\Delta^2}{\mu_8^2}\ln\frac{\sqrt{x^2+\Delta^2}-x-\mu_8}
{\sqrt{x^2+\Delta^2}-x},
\\
\int \frac{dx(2x^2+\Delta^2+x\mu_8)}{\sqrt{x^2+\Delta^2}
(\Delta^2 - 2 x \mu_8 - \mu_8^2)} &=& 
-\frac{\sqrt{x^2+\Delta^2}}{\mu_8}
+\frac{\Delta^2}{2\mu_8^2}\ln
\frac{(\sqrt{x^2+\Delta^2}+x+\mu_8)(\sqrt{x^2+\Delta^2}-x)}
{(\sqrt{x^2+\Delta^2}-x-\mu_8)(\sqrt{x^2+\Delta^2}+x)},
\\
\int \frac{dx(\sqrt{x^2+\Delta^2}-x)}{\sqrt{x^2+\Delta^2}
(\sqrt{x^2+\Delta^2}+x+\mu_8)} &=& -\frac{\sqrt{x^2+\Delta^2}-x}{\mu_8}
+\frac{\Delta^2}{\mu_8^2}\ln\frac{\sqrt{x^2+\Delta^2}+x+\mu_8}
{\sqrt{x^2+\Delta^2}+x}.
\end{eqnarray}
The other expression comes from integrating the second line 
in Eq.~(\ref{cccc}). The result is 
\begin{eqnarray}
&& 
\theta(\delta\mu-\Delta)
\int_{\mu^-}^{\mu^+}k^2dk
\frac{(E_{\Delta,k}^{-})^2+(k-\bar\mu)(k-\bar\mu_b)}
{E_{\Delta,k}^{-}\left[(E_{\Delta,k}^{-})^2-(k-\bar\mu_b)^2\right]}
\nonumber\\
&=&\bar\mu^2 \theta(\delta\mu-\Delta)
\int_{-\sqrt{\delta\mu^2-\Delta^2}}^{\sqrt{\delta\mu^2-\Delta^2}}
\frac{dx(2x^2+\Delta^2+x\mu_8)}{\sqrt{x^2+\Delta^2}
(\Delta^2-2x\mu_8-\mu_8^2)} \nonumber\\
&=& \frac{\bar\mu^2 \Delta^2}
{2\mu_8^2} \ln
\frac{\Delta^4-\mu_8^2(\delta\mu-\sqrt{\delta\mu^2-\Delta^2})^2}
{\Delta^4-\mu_8^2(\delta\mu+\sqrt{\delta\mu^2-\Delta^2})^2}\theta(\delta\mu-\Delta).
\end{eqnarray}

\end{document}